# Signatures of an Earth-class Planet in the Distant Solar System

Patryk Sofia Lykawka[1,2,3], Pedro Bernardinelli[4,5], Jonathan Horner[3]

[1] Kindai University, Faculty of Applied Sociology, Shinkamikosaka 228-3, Higashiosaka, Osaka, 577-0813, Japan

[2] National Astronomical Observatory of Japan, Center for Computational Astrophysics, Osawa 2-21-1, Mitaka, Tokyo 181-8588, Japan

[3] University of Southern Queensland, Centre for Astrophysics, Toowoomba, Queensland 4350, Australia

[4] DiRAC Institute, Department of Astronomy, University of Washington, 3910 15th Avenue Northeast, Seattle, WA 98195, USA

[5] Departamento de Astronomia, Instituto de Astronomia, Geofísica e Ciências Atmosféricas da USP, Cidade Universitária, 05508-900 São Paulo, SP, Brazil

## Abstract

Edgeworth-Kuiper belt formation models fail to explain the architecture of the Solar system beyond 50 au, leaving the origins of detached objects (perihelion $q$ > 40 au), high-inclination populations (inclination $i$ > 45°), and extreme trans-Neptunian objects a persistent mystery. To resolve these anomalies, we performed comprehensive 4.5 Gyr N-body simulations incorporating grainy Neptune migration alongside two scenarios: primordial scattering planets (Mars- to Earth-mass embryos) and a distant, stable Earth-class planet with 1–2 $M_\oplus$ (hereafter 'Planet X2'). The primordial scattering planet scenario failed to reproduce the observed distant populations in the modern Solar system because the embryos were typically ejected within a few to tens of megayears, yielding results similar to our control models without extra planets. In contrast, a 2-$M_\oplus$ Planet X2 orbiting at a semimajor axis $a$ = 350–500 au ($q$ = 200 au, $i$ = 1°–15°) generates a substantial detached population—including a crucial low-inclination component beyond ~150–200 au, creates a robust high-inclination reservoir, and replicates the orbital niches of all known extreme TNOs, such as Sedna and 2012 VP$_{113}$. The model's distant populations match all intrinsic population ratios and bulk orbital properties. Additionally, the model successfully accounts for inner TNO populations within ~100 au. An evaluation using the Dark Energy Survey Simulator demonstrates that our preferred models satisfy observational distributions for detected inclination and perihelion populations, although a discrepancy in the semimajor axis distribution highlights an area for future refinement. Overall, these findings provide evidence for an undiscovered planet and offer testable predictions to guide future surveys, such as the Vera C. Rubin Observatory's LSST.

## 1. Introduction

Over several decades of past and ongoing observational surveys, a vast population of trans-Neptunian objects (TNOs) with semimajor axes $a > 30$ au has become well-documented and characterised (e.g., Jewitt et al. 1998; Trujillo et al. 2001; Schwamb et al. 2010; Petit et al. 2011, 2017; Sheppard et al. 2011, 2016; Adams et al. 2014; Bannister et al. 2018; Bernardinelli et al. 2022). These TNOs comprise the primordial remnants of planet formation in the outer Solar system, preserving vital evidence of giant planet migration, the structure of the early protoplanetary disk, and the origins of diverse minor body populations (e.g., Malhotra 1993; Fernandez & Ip 1996; Hahn & Malhotra 1999; Schulz 2002; Morbidelli et al. 2008; Lykawka 2012; Dones et al. 2015; Nesvorný 2018; Horner et al. 2020; Prialnik et al. 2020; Gladman & Volk 2021; Kaib & Volk 2024). Consequently, the physical properties and orbital distributions of TNOs provide crucial constraints for modelling the evolution of the Solar system. In particular, any robust formation model for the trans-Neptunian region must consistently account for the distinct characteristics of the major TNO populations (Lykawka & Mukai 2007b; Gladman et al. 2008). While these constraints are detailed in our recent work (Lykawka & Ito 2023; Lykawka et al. 2026, hereafter Paper 1), a summary is provided below.

*Resonant TNOs* are trapped in various mean motion resonances (MMRs) with Neptune. Substantial resonant populations have been identified in the 1:1, 5:4, 4:3, 3:2, 5:3, 7:4, and 2:1 resonances, all located within 50 au. Additionally, an increasing number of objects have been found trapped in more distant n:1, n:2, and other MMRs located beyond 50 au (e.g., Gladman et al. 2008; Gladman et al. 2012; Volk et al. 2016; Bannister et al. 2018; Volk et al. 2018; Bernardinelli et al. 2022; Crompvoets et al. 2022). Although surveys can detect objects currently occupying resonant regions, determining whether a given object is truly Gyr-stable or merely experiencing transient resonance sticking (persisting for only a few million years) requires N-body integrations (e.g., Nesvorny & Dones 2002; Lykawka & Mukai 2007a; Lykawka & Mukai 2008; Li et al. 2014; Yu et al. 2018; Li et al. 2020; Crompvoets et al. 2022; Lykawka & Ito 2023; Porter et al. 2024; Li 2026).

*Classical TNOs* within the classical belt ($a \approx 40$–$50$ au) bifurcate into two distinct groups: the 'cold' population, characterised by low inclinations ($i < 5°$) and eccentricities ($e \lesssim 0.1$), and the 'hot' population, which exhibits higher inclinations and larger eccentricities ($i > 5°$, $e \lesssim 0.25$; e.g., Gladman et al. 2008; Nesvorny et al. 2020)[1]. Their divergent orbital characteristics along with distinct physical properties (e.g., surface colors; see details in Gladman & Volk 2021) imply disparate evolutionary paths. Specifically, while the cold classicals likely formed *in situ* beyond ~30 au, the hot classicals originated in more interior regions before being scattered outward during Neptune's migration (Gomes 2003; Nesvorný et al. 2020; Gladman & Volk 2021; Bernardinelli et al. 2025).

*Scattered TNOs* follow high-eccentricity orbits with perihelion distances $q \lesssim 40$ au, a signature of ongoing or past gravitational interactions with Neptune. Conversely, *detached TNOs* occupy similarly eccentric orbits but maintain larger perihelia ($q \gtrsim 40$ au), effectively decoupled from Neptune's influence over Gyr timescales (Duncan & Levison 1997; Gladman et al. 2002; Lykawka & Mukai 2007b; Batygin et al. 2021). These flat perihelion cuts are appropriate for characterising the representative cohorts at $a < 250$ au, where almost all known TNOs reside. Beyond this range, the same threshold is maintained for simplicity and to ensure consistency with previous work[2]. Within the distant trans-Neptunian region ($a$ = 50–1500 au), the estimated intrinsic ratio of scattered to detached objects is $\lesssim$ 1 (Lykawka & Ito 2023; Beaudoin et al. 2023; Bernardinelli et al. 2025). As discussed in Paper 1, canonical mechanisms such as resonances, von Zeipel-Lidov-Kozai (vZLK) effects, chaotic diffusion, or Neptunian migration fail to reproduce the observed detached population. This is particularly true for low-inclination ($i$ < 20–25°) or highly distant detached members ($a$ > 150–250 au). Consequently, an additional excitation mechanism beyond the known planets is likely required to account for their origin and current abundance, such as perturbations from distant massive bodies (Gladman & Chan 2006; Lykawka & Mukai 2008; Batygin & Brown 2016; Kaib et al. 2019; Huang et al. 2022b; Lykawka & Ito 2023) or early close stellar flybys (Nesvorný et al. 2023; Pfalzner et al. 2024).

[1] Although a more precise dynamical classification relies on proper inclinations (Van Laerhoven et al. 2019; Gladman & Volk 2021), the difference between the ecliptic and proper inclination distributions of the classical populations is minor (Huang et al. 2022a). Therefore, for simplicity, and given the primary focus of this work on distant populations beyond the classical belt (> 50 au), using ecliptic inclinations is acceptable.

*High-i TNOs* ($i$ > 45°) constitute a puzzling subpopulation that giant-planet instability (henceforth 'instability') and Neptune-migration models cannot adequately explain, even when those models include jumping and grainy migration scenarios (e.g., Lykawka & Mukai 2007b; Nesvorný & Vokrouhlický 2016; Nesvorný et al. 2020; Gladman & Volk 2021; Kaib et al. 2025; Paper 1). While these models predict that fewer than ~1–3% of distant TNOs should intrinsically reach $i$ > 45°, the *apparent* fraction of observed distant TNOs exhibiting such high inclinations is ~2–3% in both the AstDyS database and the Dark Energy Survey catalogues. Because observational biases heavily suppress the detection of objects with higher inclinations, these apparent figures must substantially underrepresent the true intrinsic population (Gladman et al. 2012; Shankman et al. 2017; Kavelaars et al. 2020; Gladman & Volk 2021). Therefore, the true *intrinsic* fraction of high-$i$ TNOs must be significantly larger than the models suggest, again necessitating an additional excitation mechanism beyond the known planets (presumably the same mechanism needed to produce the detached population). In this work, we conservatively adopted the apparent fractions derived from current observational catalogues (2% for the bulk distant population, and 1% and 7% within the scattered and detached subpopulations, respectively) as lower-limit benchmarks for our simulated intrinsic high-$i$ populations.

*Extreme TNOs* are prominent members of the detached or high-$i$ populations, characterised by unusually distant perihelia (e.g., Sedna, $q \approx 76$ au) and/or highly inclined, peculiar orbits (e.g., 2014 $YX_{91}$, $i \approx 63°$) (Brown et al. 2004; Sheppard et al. 2019; Bernardinelli et al. 2022; Chen et al. 2025). Their existence poses a fundamental challenge to any model of the trans-Neptunian region, as vZLK interactions (e.g., Ito & Ohtsuka 2019) within Neptune's MMRs during and after its migration cannot reproduce them (Lykawka 2012; Sheppard et al. 2016; Lykawka & Ito 2023). Accordingly, the extreme population is categorised into four dynamical regimes: $q$ > 60 au, $i$ > 60°, $q$ > 50 au with $i$ < 20°, and $i$ > 50° with $q$ < 40 au. Standard migration models fail to populate these specific regions of orbital space (Paper 1), suggesting that alternative dynamical mechanisms are required to explain the orbits of these exceptional objects.

The resonant, classical, and scattered populations provide essential clues regarding giant-planet evolution and the dynamical history beyond 30 au. However, the detached, high-$i$, and extreme populations offer the most sensitive constraints for modelling the formation of the outer Solar system. It is therefore vital that ongoing and upcoming wide-field surveys fully characterise these populations (Bannister et al. 2018; Bernardinelli et al. 2022; Trilling et al. 2024; Kurlander et al. 2025). Figures 1 and 2 illustrate the orbital structure of known TNOs within the inner and distant trans-Neptunian regions, respectively.

[2] From a phase-space perspective, the threshold separating chaotic diffusion from 'truly' decoupled regimes shifts towards larger perihelia for more distant orbits with $a \gtrsim 250$ au (Hadden & Tremaine 2024; Belyakov & Batygin 2026). Nevertheless, for classification purposes, implementing a variable perihelion cut at such distances provides a negligible statistical advantage, especially considering the paucity of detections and the associated uncertainties in population estimates. Furthermore, particles dynamically scattered outward by the giant planets from a protoplanetary disc within 30 au or a primordial scattered disk ($q$ < 35 au) retain perihelia within ~40 au at $a \gtrsim$ 200–250 au and $a \lesssim$ 1000–1500 au, provided there are no perturbations from distant planets or stellar flybys (Gladman et al. 2002; Gomes et al. 2005; Lykawka & Mukai 2007c; Kaib & Volk 2024; Nesvorny et al. 2023; Lykawka & Ito 2023; Paper 1). Therefore, a flat cut defining $q \lesssim 40$ au is physically consistent with tracing both the active and past scattering reservoirs and, conversely, $q \gtrsim 40$ au with the detached reservoir.

As outlined in Paper 1, a broad consensus suggests that the early outer Solar system initially comprised five or six giant planets in a compact orbital configuration, surrounded by a massive primordial disk. The system subsequently underwent a dynamical instability, followed by a phase of residual, slow migration of the surviving giant planets driven by the scattering of massive disk bodies (Fernandez & Ip 1984; Nesvorný 2011, 2018; Nesvorny & Morbidelli 2012; Clement et al. 2021). During its outward migration, Neptune profoundly sculpted the trans-Neptunian region by scattering objects and sweeping multiple MMRs through the population. This process is considered the primary mechanism for capturing stable resonant TNOs (Malhotra 1995; Hahn & Malhotra 2005; Pike & Lawler 2017). Notably, Neptune's migration likely exhibited 'grainy' behaviour, characterised by variations in its semimajor axis caused by encounters with hundreds or thousands of Pluto-class primordial bodies (objects with masses comparable to those of Pluto and Eris) and potentially a few larger embryos (e.g., Nesvorný & Vokrouhlický 2016; Shannon & Dawson 2018; Lawler et al. 2019; Kaib et al. 2024; Kaib et al. 2025). This graininess and the direct perturbations from these bodies reduced the efficiency of resonance capture and long-term retention, providing a likely explanation for the relatively small observed resonant populations (Paper 1). While the hot classical and scattered populations (along with a minor fraction of detached TNOs) formed through this combination of scattering and resonant interactions, the cold classical population likely formed *in situ* prior to migration. The low eccentricity, low inclination, and high binary fraction of cold classicals imply that they remained largely unperturbed by the giant planets (e.g., Noll et al. 2008; Batygin et al. 2011; Dawson & Murray-Clay 2012; Wolff et al. 2012). Consequently, the instability was likely modest, with Neptune's eccentricity peaking between 0.05 and 0.1 before rapid damping (Dawson & Murray-Clay 2012; Nesvorný 2021). Under this scenario, post-instability migration acted as the dominant driver in sculpting the region beyond ~20 au. Finally, as discussed earlier, the bulk of the detached, high-*i*, and extreme TNO populations likely formed via mechanisms other than perturbations from the known giant planets and Pluto-class bodies.

In Paper 1, we demonstrated that Neptune's grainy migration can reproduce the primary features of the inner trans-Neptunian region within ~50 au, including the resonant, classical, and scattered populations. However, all models tested in that work severely underproduced the detached, high-*i*, and extreme populations, and failed entirely to generate key observed subpopulations, such as detached TNOs with $a$ > 250 au or $i$ < 20°. Early stellar flybys or an undetected distant planet could potentially resolve these discrepancies. Several studies have suggested that close stellar flybys in the Sun's birth cluster can shape the radial structure of TNOs beyond ~50–60 au (e.g., Kenyon & Bromley 2004; Kobayashi et al. 2005; Punzo et al. 2014; Nesvorný et al. 2023; Pfalzner et al. 2024). However, current dynamical constraints derived from the orbital properties of TNOs beyond ~200 au indicate that the likelihood of such close encounters remains tightly restricted (Hu et al. 2025). Other studies have proposed that an as yet undiscovered planet is the primary driver dynamically sculpting the distant trans-Neptunian region over Gyr timescales (e.g., Lykawka & Mukai 2008; Batygin & Brown 2016). Stable resonant TNOs provide a crucial constraint in this scenario. While these objects can survive for the age of the Solar system by remaining shielded from Neptune, gravitational perturbations from a hypothetical distant planet can destabilise them, eventually transforming them into scattered or detached objects (Lykawka & Mukai 2008; Clement & Sheppard 2021; Lykawka & Ito 2023). Because long-lived (>1 Gyr) populations exist in several

MMRs, high-eccentricity resonant members impose stringent constraints on the mass and orbit of any hypothetical outer planet. Lykawka & Mukai (2008) demonstrated that preserving the known stable 3:2, 2:1, and 5:2 resonant populations requires such a planet to have $q > 80$ au. Recently, Lykawka & Ito (2023) updated this limit to $q > 145$ au (for less massive planets with $m = 0.5–1\ M_\oplus$, where $M_\oplus$ is the Earth mass) and $q > 195$ au (for $m = 2–3\ M_\oplus$).

The hypothesis that an undiscovered distant planet may be dynamically sculpting the trans-Neptunian region dates back to the 1990s, when a Mars-class body on a close-in orbit within 50–100 au was proposed to explain the orbital architecture of the classical belt and the rapid decline in TNO density beyond $a \approx 50$ au (for details on this 'Planet X' scenario, see Lykawka & Mukai 2008). After accounting for multiple constraints within the trans-Neptunian region and utilising updated observational samples, subsequent improved models suggested that an Earth-class body (i.e., an object with a mass in the range of 0.5–2 $M_\oplus$) may remain undetected beyond ~100 au (Lykawka & Mukai 2008; Lykawka & Ito 2023). While distinct in its objectives and predicted orbit-mass properties compared to the canonical model, we designate this refined scenario as 'Planet X2' to reflect the foundational idea of an undiscovered planet in the trans-Neptunian region. This corresponds to the 'Kuiper Belt planet' scenario described in Lykawka & Ito (2023). The primary motivation for this scenario is that the long-term gravitational influence of Planet X2 could account for several observed orbital properties and peculiarities among TNO populations (Lykawka & Mukai 2008; Lawler et al. 2017; Lykawka & Ito 2023; Siraj et al. 2025b). For example, Lykawka & Mukai (2008) demonstrated that an outer planet migrating to a final orbit of $a = 100–250$ au, with $q > 80$ au and $i = 20–40°$, could simultaneously reproduce the four main TNO populations as they were constrained at that time.

Separately, the apparent clustering of orbital angles among a subset of distant TNOs beyond 150–250 au motivated the Planet 9 (P9) hypothesis, which proposes a super-Earth with 5–10 $M_\oplus$ orbiting at ~380–500 au ($q \approx 300–400$ au) as the cause of this feature (Trujillo & Sheppard 2014; Batygin & Brown 2016). Along similar lines, a closer-in 'mini-P9' version with 4.4 $M_\oplus$ and $a = 290$ au ($q = 200$ au) has also been proposed (Siraj et al. 2025a). Although independent studies have contested the statistical significance of the alleged orbital clustering (Lawler et al. 2017; Shankman et al. 2017; Bernardinelli et al. 2020; Gladman & Volk 2021; Napier et al. 2021), attributing it to observational biases and the small sample size of the currently known subset of distant TNOs, some subsequent works suggest that P9 could generate the detached and highly inclined populations, as well as a low-inclination scattered population, all of which are potentially compatible with observations (Batygin et al. 2019; Batygin et al. 2024). On the other hand, Kaib et al. (2019) found that the P9 configurations favoured in these studies overproduce high-inclination objects relative to the well-charactersed Outer Solar System Origins Survey (OSSOS) observational sample. Moreover, the orbits and long-term evolution of recently discovered distant TNOs appear inconsistent with P9 predictions, including the mini-P9 scenario (Chen et al. 2025; Cheng et al. 2026; Siraj et al. 2026).

More recently, Lykawka & Ito (2023) refined the Planet X2 scenario using extensive simulations that incorporated Mars- to Earth-class bodies and updated observational constraints. By comparing their results with the OSSOS dataset (Bannister et al. 2018; Lawler et al. 2018a), they demonstrated that an undetected Earth-class planet on an orbit with $a = 250–500$ au, $q \approx 200$ au, and $i = 30°$ can consistently reproduce the resonant, scattered, detached, high-$i$, and extreme TNO populations while preserving known stable resonant objects. However, the Lykawka & Ito (2023) model was developed without giant-planet migration and omitted the inner trans-Neptunian region ($a < 50$ au). To address these limitations, in this study we expand upon that work by investigating the Planet X2 scenario within a comprehensive, physically motivated grainy migration model. Furthermore, we extend these simulations by including a massive disk populated by several thousand planetesimals and primordial Pluto-class bodies, while accounting for the entire inner trans-Neptunian region. Finally, under this framework, we examine the trans-Neptunian orbital structure and compare our results with the well-characterised Dark Energy Survey (DES) observational sample, which covers all major TNO classes and features a more extensively sampled distant population.

## 2. Methods

The numerical methods employed in this work are consistent with those of Paper 1. For completeness, we summarise the baseline simulation setup below and subsequently describe the new simulations performed for this study that extend the original framework. We refer the reader to Paper 1 for a more comprehensive description.

### 2.1 Grainy migration model

In Paper 1, we conducted baseline N-body simulations to investigate the formation and dynamical evolution of the trans-Neptunian region over the age of the Solar System in the absence of additional planets. These control simulations included Jupiter, Saturn, Uranus, and Neptune, with initial semimajor axes of 5.26 au, 9.2 au, 18.7 au, and 24.0 au, respectively[3]. The models considered grainy migration of the giant planets driven by interactions with a massive population of planetesimals and 1500 Pluto-class bodies within 30 au in the protoplanetary disk (Nesvorný & Vokrouhlický 2016; Kaib et al. 2024). All simulations incorporated a two-part disk consisting of an interior region (25–30 au swept by Neptune during its outward migration) and an exterior region (30–49 au representing the primordial cold classical belt). The interior disk comprised both massive and massless components. To ensure that Neptune reached a final orbit near its current position after 4.5 Gyr of evolution, the massive interior disk was initialised with a total mass of ~14–14.5 $M_\oplus$ (e.g., Nesvorný 2018). This massive component comprised 1500 Pluto-class bodies (each with a mass of $3.333\times10^{-3}$ $M_\oplus$, accounting for 5 $M_\oplus$ in total) and 13,500 (or 14,250) smaller massive planetesimals (each with a mass of $6.667\times10^{-4}$ $M_\oplus$, accounting for the remaining 9.0–9.5 $M_\oplus$). By default, all planets interacted gravitationally with each other and with the disk of small and Pluto-class bodies, but these disk objects did not interact with each other. To improve surviving particle statistics, the interior disk also included ~87,000–139,000 test particles, while the exterior disk was represented by a dynamically cold population of 2500 test particles. The initial eccentricities and inclinations for the interior (exterior) disk were set to 0.2 (0.01) and 5° (0.5°), respectively. Overall, these initial conditions are broadly consistent with those employed in several previous models (e.g., Nesvorný 2015a; Nesvorný & Vokrouhlický 2016; Kaib & Sheppard 2016; Volk & Malhotra 2019; Nesvorný et al. 2020). In these simulations, the giant planets migrated gradually to their current orbits through gravitational interactions with the disk objects.

To isolate the specific effects of giant planets on the final architecture after 4.5 Gyr of dynamical evolution, we aggregated the systems from the '4GP', '4GP_jNep', and '4GP-pluSG' models described in Paper 1 into a single ensemble of 50 systems. This composite sample serves as our control grainy migration model (Figure 3). We combined these three models because they yielded remarkably similar results in the distant trans-Neptunian region, which is the primary focus of this work. Furthermore, the results of this control model agree well with those of several models published in the literature (see Lykawka & Ito 2023 and Paper 1 for details). Consequently, the control model is treated as representative of the current consensus.

---

[3] The initial eccentricities of the four giant planets were 0.07, 0.07, 0.01, and 0.01, while their respective inclinations were 2°, 2.5°, 1.5°, and 1.5°.

2.2 Planetary perturber model

Building on our baseline framework, we conducted additional N-body simulations to investigate the effects of adding other planetary-mass bodies to the system during and after giant planet migration (Section 2.3). In these simulations, we included either 'scattering' or 'outer' planets in all systems. Scattering planets (the Emb models) were placed within 30 au on Neptune-crossing orbits to represent primordial Mars- to Earth-mass embryos formed within the interior disk. An outer planet refers to any hypothetical undiscovered planet located beyond 100 au on a stable orbit; here, we focus on Planet X2 as the representative of the outer planet scenario for consistency with our previous work (Section 1). We modelled Planet X2 by placing it on orbits with $q \geq 200$ au, $a = 250$–$500$ au, and $i = 1$–$30°$ (the PX2 models).

Guided by our previous results regarding the Planet X2 scenario, we varied the masses and orbits of the hypothetical outer planets across different models while maintaining consistency with the observed stable resonant populations (Lykawka & Ito 2023). In this study, we extended the Planet X2 scenario by incorporating the grainy migration of the giant planets, utilising a high-resolution interior disk, and including an exterior disk to represent the primordial classical belt. We then assessed the influence of Planet X2 on the trans-Neptunian region during both the migration and post-migration stages of giant-planet evolution. For completeness, we also independently examined the P9 scenario by incorporating a hypothetical perturber representing Planet 9. For this model, we adopted the updated parameters from Brown & Batygin (2021), specifically a 6.2 $M_\oplus$ super-Earth with $a = 380$ au, $q = 300$ au, and $i = 16°$. Table 1 summarizes the initial conditions for all models considered in this study.

2.3 New N-body simulations

We investigated the long-term dynamical evolution of all systems in our models by executing simulations for a total duration of 4.5 Gyr. These numerical calculations were conducted using a modified version of the MERCURY integrator (Chambers 1999; Hahn & Malhotra 2005) with a time-step of 0.6 yr. Objects were removed from the simulations if their heliocentric distances fell below 3 au or exceeded 1000 au (or 2400 au in models incorporating outer planets), following previous outer Solar system studies (e.g., Kaib & Sheppard 2016; Lykawka & Ito 2023; Kaib et al. 2025). The dynamical effects of galactic tides and stellar flybys were omitted from our simulations as they become significant only in the inner Oort Cloud region at $a > 1500$–$2000$ au (Gladman et al. 2008; Trujillo & Sheppard 2014; Nesvorný et al. 2017; Saillenfest 2020; Kaib & Volk 2024). In addition, this work aimed to investigate the gravitational effects of additional planets within our region of interest, as defined by these heliocentric cut-offs. All simulation runs were executed on single CPU cores, with most individual runs requiring approximately 4–5 months to complete. At 4.5 Gyr, Neptune analogues reached final orbits within the range of 29–31 au. Consistent with the protocol established in Paper 1, we normalised object orbits using the formula $30.1 \times a_{\rm run} / a_{\rm Nep,run}$ (where $a_{\rm run}$ and $a_{\rm Nep,run}$ represent the final semimajor axes of the object and Neptune, respectively) and rescaled the perihelia accordingly. To ensure a representative analysis, we selected the successful systems (runs) that produced orbital architectures consistent with the four main TNO populations and co-added these results for each model. This approach enabled us to probe individual models at high resolution and statistically summarise the results, thereby mitigating the influence of stochastic variations inherent in individual systems (Volk & Malhotra 2019). Furthermore, in the analysis of these systems, the protoplanetary disk was assumed to be confined within 47 au. This confinement is required to explain the primary properties and population ratios of TNOs, while satisfying other key constraints in the inner trans-Neptunian region (see Paper 1 and references therein).

Finally, following the characterisation of the trans-Neptunian populations described in Section 1, we classified the simulated populations using the following scheme:

(1) *Resonant* (stable) objects: bodies that exhibited libration for longer than 2 Gyr and up to 4.5 Gyr. We employed the RESTICK code to track the evolution of resonant arguments across several potential MMRs without restricting the analysis to preselected resonances. The code identifies resonant motion by continuously evaluating MMRs with orders up to 100 located near each object's semimajor axis during its orbital evolution (Lykawka & Mukai 2007a, Lykawka & Mukai 2007c);

(2) *Classical* objects: non-resonant bodies located within the classical belt at $a$ = 40–50 au and $q >$ 37 au, with cold and hot components distinguished by inclinations of $i < 5°$ and $i \geq 5°$, respectively;

(3) *Scattered* objects: non-classical bodies with $25 < q < 40$ au;

(4) *Detached* objects: non-classical bodies with $q > 40$ au;

(5) *Centaurs*: bodies with $q < 25$ au;

(6) *High-i* objects: bodies with inclinations $i > 45°$; and

(7) *Extreme* objects: bodies defined by one of specific conditions: $q > 60$ au, $i > 60°$, $q > 50$ au with $i < 20°$, or $i > 50°$ with $q < 40$ au.

Lastly, after obtaining the final systems for selected models, we utilised the DES Survey Simulator to compare the results with the well-characterised observational sample (Section 4; Bernardinelli et al. 2022; Bernardinelli et al. 2025). A survey simulator evaluates the detectability of model objects under observing conditions, thereby imparting the appropriate observational biases to the resulting systems. These biased populations can then be compared directly with actual observational data. We favoured the DES dataset over the OSSOS dataset because it provides a larger sample of objects with $a > 50$ au, as well as a higher number of TNOs on detached and highly inclined orbits. These specific orbital niches represent the primary populations of interest in this work. However, owing to the currently limited sample size of such TNOs, discriminating between models based solely on survey simulator evaluations remains challenging (e.g., Lawler et al. 2017). Consequently, only models that successfully satisfied the fundamental intrinsic constraints discussed in Section 1 were selected for evaluation with the DES Survey Simulator, with overall model success requiring reasonable compatibility with DES observations once observational biases were applied.

## 3. Results

In general, irrespective of the presence of scattering planets or Planet X2, interactions with disk objects primarily drove Neptune's orbital evolution. Under this framework, Neptune migrated outward and swept through the protoplanetary disk before gradually settling into its final orbit, with the majority of its migration completed within the first 1 Gyr. Following the approach adopted in Paper 1, we excluded from the analysis any model systems lacking resonant populations within 50 au or exhibiting excessive stirring of the cold component of the classical belt, as these outcomes directly contradict the observational constraints provided by the modern Solar system. In these excluded systems, given the low eccentricities ($e < 0.1$) and non-instability configurations of Jupiter and Saturn (as was also the case for non-excluded systems), Uranus–Neptune interactions, scattering planets, or both played the primary role in overexciting the cold classical belt and depleting major resonances (e.g., the 3:2 and 2:1 MMRs). Specifically, this excitation was driven either by stochastic Uranus–Neptune perturbations involving their mutual 2:1 resonance or by scattering planets persisting in the system for more than several tens of Myr (including stranded ones within 50 au). These results are consistent with Kaib et al. (2024) and Graham & Volk (2024). Table 1 lists the details of the models explored in this study, along with the number of final systems (runs) utilised in the analysis, while Table 2 summarises the orbital properties of the final systems obtained from these models.

Overall, the final systems reproduced the major dynamical TNO populations and were retained for further analysis. We considered all objects across these systems that survived with $q > 25$ au after 4.5 Gyr of evolution. In the following sections, we analyse the final orbital configurations across three scenarios: the dynamical influence of scattering planets (Section 3.1), gravitational sculpting by Planet X2 (Section 3.2), and a comparison with the Planet 9 scenario (Section 3.3).

### 3.1 Scattering planets

Our results demonstrate that planets embedded within the interior disk (< 30 au) evolve stochastically as they interact with disk objects and undergo gravitational scattering by Neptune and the other giant planets. A substantial fraction of these scattering planets excessively excited the cold classical belt through their perturbations, either by acquiring stable orbits within ~50 au (i.e., becoming stranded) or by remaining for overly long periods within the system (Section 3.1.1). After accounting for these effects alongside Uranus–Neptune perturbations, we found that the failure rates were 22.5% (9/40), 47.5% (19/40), 50% (20/40), 60% (24/40), and 67.5% (27/40) for the five models of scattering planets tested in this work: a single 3x-Mars-mass planet (model Emb-1M3), a single 5x-Mars-mass planet (model Emb-1M5), a single Earth-mass planet (model Emb-1E), ten Mars-mass planets (model Emb-10M), and two Earth-mass planets (model Emb-2E) scattering planets (Table 1). In other words, of the 200 individual systems simulated for this scenario, 49.5% (99/200) produced planetary systems that were markedly inconsistent with the current Solar system. These results also indicate that the probability of failure increases with the scattering planet mass or the total number of scattering planets within 30 au in the early Solar system. The following subsections focus on the remaining systems that met the selection criteria established in Section 3.

#### 3.1.1 Orbital evolution, lifetimes, and dynamical influence

Across all systems included in our final sample, no scattering planets survived to the end of the 4.5 Gyr simulations. In each case, these planets were short-lived and eventually lost from the system via gravitational ejection by Jupiter (upon reaching heliocentric distances > 1000 au) or through collisions with a giant planet or the Sun. The median dynamical lifetimes of individual scattering planets across models Emb-1M3, Emb-1M5, Emb-1E, Emb-10M, and Emb-2E were approximately 9 Myr, 9 Myr, 6 Myr, 4–38 Myr, and 8 Myr, respectively. In agreement with the findings of Kaib et al. (2025), these results imply that single scattering planets that avoid excessively exciting the inner trans-Neptunian region tend to spend ~10 Myr interacting strongly with Neptune on scattering orbits ($q < 35$ au) before being ejected. These shorter lifetimes, relative to the ~100–200 Myr survival timescale of the 2 $M_\oplus$ rogue planet in Huang et al. (2022b), arise because our embryos experience unstable, planet-crossing orbits during giant planet migration, rather than being initialised on wider, decoupled orbits ($a = 300$ au, $q = 40$ au, $i = 20°$ in that work). In model Emb-10M, mutual interactions among the multiple Mars-mass planets extended lifetimes into the tens of Myr, with some individual bodies persisting beyond 100 Myr. Despite this longevity, the dynamical perturbations exerted by these planets in the inner region remained minor overall. These results complement those of Lykawka & Ito (2023), who found that hypothetical Mars-class planets exert only minor effects on the orbital structure beyond 50 au over 4.5 Gyr of dynamical evolution.

Overall, the short dynamical lifetimes of the scattering planets in the final systems ensure that the trans-Neptunian orbital structure remains closely similar to the control model. This consistency is reflected in the typical distributions across *q–i*, *a–i*, and *a–q* orbital space (Figures 3–5). The resonant, classical, scattered, and detached populations also closely mirror those of the control case, confirming that these transient planets typically leave only minor dynamical signatures in the distribution of TNOs after 4.5 Gyr. Specifically, classifying objects into the major TNO dynamical categories enables a direct comparison with the outcomes of the control model (indicated in parentheses below). Across all simulations featuring scattering planets, the ratio of the 3:2 to 2:1 resonant populations fell in the range of 1.0–1.4 (1.7 ± 0.6), whereas the ratio of the 3:2 to 5:2 resonant populations spanned 9.5–26.5 (26.8 ± 30.7). In total, 28.9–40.3% (30.0 ± 16.1%) of all surviving objects were trapped in resonances at the end of the simulations. Finally, the ratio of scattered to detached populations was 3.9–5.6 (3.7 ± 1.2), and high-*i* objects accounted for 1.7–2.7% (3.0 ± 1.4%) of the final population after 4.5 Gyr. Furthermore, the probability of generating objects on extreme orbits remained negligible across all these models. The high degree of similarity in orbital structure beyond 50 au compared to the control model confirms that scattering planets are inefficient at producing detached, high-*i*, or extreme objects. Consequently, the presence of scattering planets alone cannot account for the observed scattered-to-detached ratio, the high-*i* population fraction, or the origin of extreme orbits.

Despite the general consistency with the control model, a detailed inspection of the final systems reveals a few differences induced by scattering planets. First, in agreement with previous studies (Huang et al. 2022b; Kaib et al. 2025), these planets facilitate the formation of detached objects within ~100 au, including those with low inclinations. However, the resulting detached population generally remains minor compared with its scattered counterpart (exceptional systems are discussed in Section 3.1.2). Second, dynamical depletion within the 3:2 MMR is more pronounced than in other MMRs, leading to reduced 3:2-to-5:2 resonant population ratios. For dynamical lifetimes of the scattering planet exceeding 20 Myr, the 3:2-to-5:2 ratio appears anticorrelated with the mass, dynamical lifetime, or both, of the planet. Specifically, we determined median ratios of 23.3, 26.5, 18.1, 9.7, and 9.5 for models Emb-1M3, Emb-1M5, Emb-1E, Emb-10M, and Emb-2E, respectively. Given that the ratio obtained in the control model (26.8) is significantly higher than the observed intrinsic (unbiased) 3:2-to-5:2 ratio of ~1.0–2.0 derived from well-characterised surveys (e.g., Gladman et al. 2012; Volk et al. 2016; Crompvoets et al. 2022; Bernardinelli et al. 2025), the reduced ratios in some models tentatively indicate that scattering planets help alleviate this discrepancy. However, given the complex secular landscape near the 3:2 MMR, subtle variations in the final giant-planet orbits might also contribute (e.g., Nesvorný & Roig 2000; Tiscareno & Malhotra 2009). Although co-adding our simulated systems likely mitigates these effects (see Section 2.3), definitively isolating the role of the scattering planets requires high-resolution and long-term tracking of resonant escape rates, which is left for future work. Third, we analysed the dynamical effects of scattering planets with longer lifetimes (10–30 Myr) on the classical belt by calculating the median eccentricity of cold classical particles within $a$ = 42.7–47.3 au (isolated from the adjacent 5:3 and 2:1 MMRs). While single sub-Earth planets matched the control model with a median of 0.06 (compiled across eight Emb-1M3 and seven Emb-1M5 systems), multiple Mars-mass as well as single/dual Earth-mass planets generated higher median eccentricities of 0.07 (nine Emb-10M systems), 0.07 (four Emb-1E systems) and 0.09 (seven Emb-2E systems), trending higher towards larger semimajor axes. This dynamical excitation may help explain the observed scarcity of low-eccentricity cold classical TNOs beyond 45 au (e.g. Lykawka & Mukai 2008; Batygin et al. 2011; Petit et al. 2011; Dawson & Murray-Clay 2012; Paper 1).

### 3.1.2 Individual systems of interest

Among the final systems used in our analysis, we identified three exceptional cases in which prolonged perturbation by scattering planets improved the model's alignment with key constraints of the distant trans-Neptunian region: systems 23, 8, and 11 from models Emb-1M3, Emb-1E, and Emb-2E, respectively. The dynamical lifetimes of the scattering planets in these systems were 413 Myr, 48 Myr, and 4 Myr/49 Myr (for the two embryos), respectively. Consistent with our earlier discussion, the primary effects of these perturbations were an enhanced abundance of scattered and detached objects extending to ~100–200 au and reduced populations in MMRs within 50 au (leading to smaller 3:2-to-5:2 resonant population ratios). These interactions yielded scattered-to-detached population ratios of 1.02, 1.01, and 2.14 beyond 50 au for these specific systems. The first two values, in particular, are compatible with observational estimates of $\lesssim 1$ (Lykawka & Ito 2023; Beaudoin et al. 2023; Bernardinelli et al. 2025).

Although these results provide a proof of concept for reproducing selected observational constraints, critical difficulties remain. First is the low likelihood of such outcomes. Each exceptional system formed out of 40 initial runs, thus representing an upper limit of approximately 2.5% probability of potential success. Furthermore, as the Emb-1M5 and Emb-10M models yielded zero such systems, the overall empirical probability across all 200 runs is roughly 1.5% (3/200). Such small probabilities stem from the high stochasticity of the evolution that one or multiple scattering planets can experience during the early Solar system. Another potential issue is the distribution of the low-*i* detached component formed in these systems, which is concentrated within 200 au, 150 au, and 100 au for systems 23, 8, and 11, respectively. These distributions are incompatible with observations according to the findings of Kaib et al. (2025), which suggest a concentration of such bodies beyond ~200 au. Furthermore, none of the three exceptional systems yielded highly inclined ($i > 45°$) populations beyond 50 au that could explain the observed apparent fraction of 2%; the specific values for these systems were 0.4%, 0.2%, and 0.2%, respectively. Consequently, when this condition is added to the full list of observables, none of the 200 runs yielded a potentially successful system. Although it remains possible that an individual system could form with properties that satisfy those constraints simultaneously if hundreds or thousands of runs were tested for a specific model, our results suggest a negligible probability of success ($\ll$ (1/200)).

#### 3.1.3 Analysis summary

We conclude that while the addition of scattering planets to the grainy migration model may help satisfy specific observational constraints, it is unlikely to account for the comprehensive orbital architecture detailed in Section 1. In particular, the extremely low inherent probability of these potentially successful systems, coupled with their inability to consistently replicate the observed high-*i* and distant detached populations, suggests that alternative dynamical mechanisms must be considered.

### 3.2 Outer planets: the Planet X2 scenario

We evaluated whether the final systems from our Planet X2 models could reproduce the distinct TNO populations and their orbital properties. Overall, these systems successfully formed the classical, resonant, scattered, and detached populations. Consequently, we analysed the collective properties of these systems and compared them with the control and scattering-planet models, as detailed below.

Our final systems produced Gyr-stable resonant populations in several MMRs, including the 1:1, 5:4, 4:3, 3:2, 5:3, 7:4, 2:1, 5:2, 3:1, 4:1, 5:1, 6:1, 7:1, 8:1, 9:1, 10:1, and 11:1 resonances. Furthermore, in agreement with earlier studies (Lykawka & Mukai 2008; Clement & Sheppard 2021; Lykawka & Ito 2023; Porter et al. 2024), the presence of Planet X2 with a sufficiently large perihelion does not preclude the retention of these modelled cohorts alongside the known stable resonant populations. Thus, our Planet X2 models successfully reproduce the stable resonant TNOs. Although the fraction of the resonant population relative to the total surviving population was ~20–40% in these models, incorporating the self-gravity of Pluto-class objects during migration modelling can reduce this fraction to values that match the observed intrinsic population[4] (Kaib et al. 2024; Paper 1). Consistent with the results of Paper 1, our models show that assuming a disk extension to 47 au yields classical belt orbital structures similar to observations. The results also confirm that the bulk of the cold and hot classical populations originated from the exterior and interior regions of the protoplanetary disk, respectively (Trujillo & Brown 2001; Petit et al. 2011; Gladman & Volk 2021). Perturbations by Neptune and Planet X2 also drove the formation and continuous dynamical interplay between the scattered and detached populations, which extended from orbits within 50 au to the outer boundaries explored in this work ($a$ ~ 1200-1400 au, consistent with the aphelion cut-off of $Q \leq 2400$ au; see Section 2). These populations are discussed in greater detail in Section 3.2.1.

The Planet X2 scenario yielded an orbital distribution in $a$–$e$–$i$–$q$ space within ~50–60 au that is in close agreement with the control model. This consistency extends to several key metrics, including resonant population ratios such as the 3:2-to-2:1 and 3:2-to-5:2 ratios. Thus, our results confirm the expectation that the perturbations of Planet X2 are negligible in the inner reaches of the Solar system. In summary, our results demonstrate that a scenario combining giant-planet grainy migration and Planet X2 is fully compatible with the primary constraints posed by the region within ~50–60 au.

We also evaluated whether any Planet X2 models could simultaneously account for the detached, high-$i$, and extreme TNO populations beyond 50 au. The following results confirm the fundamental findings of our previous studies (Lykawka & Mukai 2008; Lykawka & Ito 2023). First, our Planet X2 models generated a distant trans-Neptunian region composed of stable resonant, scattered, and detached populations, with the latter two comprising >90% of the total (including transient resonant objects). As illustrated in Figures 6–15, the resulting orbital distributions of these distant populations diverge sharply from those of the control and scattering-planet models (compare with Figures 3–5). These distinctive distributions in $q$–$i$, $a$–$i$, and $a$–$q$ space serve as unique observational signatures predicted by the Planet X2 scenario.

#### 3.2.1 Detached populations

Overall, Planet X2 drives the formation of a prominent detached population via secular perturbations that raise perihelia and decouple orbits from Neptune. Populating this reservoir in sufficient numbers requires timescales exceeding ~1 Gyr for Earth-class planets, as evidenced by the sparse detached populations produced in shorter integrations (Lykawka & Mukai 2008). Given the assumed perturber perihelion of $q \geq 200$ au in our models, we expect that the bulk of the detached population at $a \lesssim 100$ au formed via the vZLK and MMR dynamics associated with Neptune during and after its grainy migration, aligning with the findings of Nesvorný et al. (2023). These are the same mechanisms identified in the control and scattering-planet models discussed in previous sections. Conversely, perturbations from Planet X2 were responsible for generating the majority of the detached population at $a \gtrsim 100$ au. While the entire detached population in the control model is confined to $q$ ~ 50 au (with some resonant members reaching ~60 au) and $a$ < 200–

[4] A rapid, single radial displacement of Neptune during its migration (the 'jumping Neptune' scenario) could further modestly reduce the resonant population fraction, as discussed in Paper 1.

250 au, the influence of Planet X2 produces significantly broader perihelion and semimajor-axis distributions. In particular, we observe an increase in detached objects with $q$ = 40–60 au at $a \sim$ 100–200 au, alongside a systematic lifting of perihelion for objects with $q$ > 60 au at $a \gtrsim$ 100 au, which can reach hundreds of au for more distant orbits. The detached population also exhibits distinctive concentrations in $a$–$q$ space that directly reflect the orbital elements of Planet X2; more eccentric planetary orbits yield detached populations with wider ranges of semimajor axes and perihelia. Another key result of the Planet X2 models is the formation of numerous distant detached objects with low inclinations, filling a wide range of orbits unreachable by the control and scattering-planet models (see Section 3.2.4 for details). The resulting detached populations typically exhibit a broader inclination distribution and pronounced concentrations in $q$–$i$ space at $q$ > 100 au. The inclination of Planet X2 strongly dictates this structure; more inclined planets produce objects with systematically higher inclinations, whereas lower Planet X2 inclinations are associated with a larger share of low-$i$ detached objects. Finally, these results demonstrate that, unlike the control and scattering-planet models, the Planet X2 models can readily produce orbits that explain the observed subpopulations of detached TNOs with $a$ > 250 au, scattered TNOs with $q$ > 37 au and $i$ < 20°, and detached TNOs with $i$ < 20°.

Furthermore, the formation of a prominent detached population in the Planet X2 models leads to scattered-to-detached population ratios typically smaller than unity, values that are several times smaller than those obtained in the control (3.7) and scattering-planet (3.9–5.6) models. Indeed, all but two of our Planet X2 models yielded ratios consistent with the observationally constrained value ($\lesssim$ 1.0; Section 1).

In summary, the results discussed above indicate that models incorporating Planet X2 generate a detached population that successfully satisfies the observationally derived intrinsic ratio of scattered to detached objects and reproduces key orbital features of known detached objects that standard four-planet migration or embryo-scattering models fail to explain.

### 3.2.2 High-i populations

The multi-Gyr perturbations of Planet X2 drove the formation of a substantial high-$i$ population. Notably, this population concentrated across a broad region beyond approximately 200 au, and its highly inclined component closely reflected the inclination of the adopted Planet X2 configurations. Furthermore, while the inclination distribution tended to cluster below 90°, several PX2 models also produced non-negligible populations of retrograde trans-Neptunian objects ($i$ > 90°). In particular, more massive planets on more distant and inclined orbits ($m \geq 2$ M$_\oplus$, $a \geq 350$ au, and $i \geq 15°$) facilitated the generation of this retrograde population (Table 2). Crucially, this subpopulation comprises a few to several per cent of the distant population and represents a natural source for Solar system retrograde objects, as they can evolve towards smaller $a$ or $q$ during dynamical interactions with Planet X2 and the giant planets. The generation of this population provides a plausible mechanism for the origin of retrograde Centaurs and TNOs (e.g., Gladman et al. 2009; Brasser et al. 2012; Chen et al. 2016; Kaib et al. 2019; Greenstreet et al. 2020; Gladman & Volk 2021; Lykawka & Ito 2023; Pfalzner et al. 2024).

Analysis of the inclination distributions revealed that our control and scattering-planet models yielded high-$i$ fractions of 3.0% and 1.7–2.7%, respectively. Although these fractions are compatible with our conservative high-$i$ constraint ($\geq$ 2%), they are likely insufficient to explain the high-$i$ population because the intrinsic (unbiased) value must substantially exceed 2% (as well as $\gg$1% and $\gg$7% for the scattered and detached populations, respectively; see Section 1 and Table 2). In this regard, we found that PX2 models incorporating a 1 M$_\oplus$ outer planet yielded relatively low high-$i$ fractions for the entire distant population (2.3%–3.2% for $i_P$ = 1° and 3.9–14.5% for $i_P$ = 30°;

Table 2). By comparison, models with a more massive 2 $M_\oplus$ planet yielded more pronounced excitation: the overall fractions rose to 2.9–16.4% for $i_P$ = 1° and 19.5%–45.6% for $i_P$ = 30° (Table 2). A similar trend was found within the individual scattered and detached populations. Overall, these results suggest that a Planet X2 more massive than 1 $M_\oplus$ is required to produce a sufficiently abundant excited population to explain the observed highly inclined TNOs.

In summary, the results discussed above indicate that several PX2 models can successfully account for the high-$i$ TNO population. These findings also support the hypothesis that a trans-Neptunian high-$i$ reservoir (e.g., Gladman & Volk 2021) acts as a source, supplying highly inclined objects throughout the entire Solar system.

#### 3.2.3 Extreme populations

The apparent fraction of extreme TNOs at $a$ > 50 au is 1.3% (11/819), based on data from the AstDyS database (as of April 2026). However, owing to severe observational biases that hinder the discovery of objects with exceptionally high inclinations, large perihelia, or both, the intrinsic, unbiased fraction is expected to be substantially higher. In our simulations, the control and scattering-planet models yielded median fractions of only 0.5% and 0.7%, respectively. Because these fractions fall below the apparent 1.3% and remain significantly smaller than the expected intrinsic value, we conclude that these models fail to account for the origin of the extreme TNO population. In contrast, all PX2 models produced significant fractions of extreme objects beyond 50 au, ranging across 12.3%–35.1% and 43.5%–67.8% for the 1 $M_\oplus$ and 2 $M_\oplus$ perturber models, respectively (Table 2). These elevated fractions may be sufficient to explain the observed extreme population when observational biases are accounted for in future studies.

In addition, a comparison of the orbital properties of the distant extreme TNOs[5] (Figure 2) with our results reveals that the control and scattering-planet models are wholly inefficient at populating extreme orbital niches at $q$ > 60 au, $i$ > 60°, $q$ > 50 au with $i$ < 20°, and $i$ > 50° with $q$ < 40 au. Furthermore, these models are incapable of reproducing the orbits of the six members with $q$ > 50 au (specifically, 2023 $KQ_{14}$, 2012 $VP_{113}$, (90377) Sedna, 2013 $SY_{99}$, 2021 $RR_{205}$, and (541132) Leleakuhonua). These findings confirm the results of our previous analyses (Lykawka & Ito 2023; Paper 1). In contrast, the PX2 models produced extreme populations with members that generally cover all of these niches. However, a close inspection of the orbits of the six extreme TNOs with $q$ > 50 au reveals that orbits with $a$ > 800 au and $i$ < 15°, similar to those of 2013 $SY_{99}$, 2021 $RR_{205}$, and Leleakuhonua, are rarely reproduced in models where Planet X2 has $a$ = 250 au, or $a$ = 350 au with $i$ = 30°. Accounting for this additional constraint, our tested models point towards a Planet X2 with $a \gtrsim 350$ au and lower inclinations ($i$ < 30°), which successfully produces an extreme population containing objects with orbits analogous to all 11 known extreme TNOs beyond 50 au. Future simulations should target intermediate-to-wide orbits (e.g., $a \approx$ 400–600 au and $i \sim$ 5–20°) to constrain this parameter space further.

#### 3.2.4 Other distant populations

Standard models of outer Solar system formation cannot produce low-$i$ detached objects (Section 1). Consistent with past studies, our control model yielded an average of only 0.5% and 3.5% of the entire detached population with $i$ < 20° and $i$ < 25°, respectively. Additionally, these low-$i$ detached objects were predominantly confined within 100 au. The final systems produced by the scattering-planet models yielded closely similar results. In contrast, the PX2 models produced a substantial number of detached objects at low inclinations. Crucially, we found that only a small fraction of these objects were located at $a < 100$ au. Indeed, the detached populations with low inclinations obtained in these models are concentrated beyond $a \sim 150$–200 au (Figures 6–15), with median semimajor axes typically exceeding several hundred au (Table 2). Therefore, the PX2 models satisfy the additional observational constraint requiring the low-$i$ detached population to be concentrated beyond ~100 au (Lykawka & Ito 2023; Kaib et al. 2025).

[5] As of April 2026, eleven extreme TNOs on distant orbits at $a$ > 50 au are known: 2015 $UO_{105}$, 2014 $YX_{91}$, 2014 $UN_{225}$, 2017 $FO_{161}$, 2023 $KQ_{14}$, 2012 $VP_{113}$, (768325) 2015 $BP_{519}$, (90377) Sedna, 2013 $SY_{99}$, 2021 $RR_{205}$, and (541132) Leleakuhonua.

The inclination distribution of distant populations provides critical insight into the primary past and ongoing dynamical mechanisms operating in the outer Solar system. Specifically, the observed lower inclinations of detached TNOs beyond ~100–200 au may not be solely an artefact of observational bias. Based on that premise, Hu et al. (2025) suggested a tentative observational constraint that Sednoids (defined at $a = 200$–1500 au and $q > 60$ au) with $i < 30°$ should constitute >50% of that population. Under this constraint, a Planet X2 with $i = 30°$ is disfavoured. Given that this intrinsic constraint crucially limits models predicting large high-$i$ fractions in distant populations (e.g., hypothetical outer planets with $i \gtrsim 30°$, or early close stellar flyby models), future observations are essential to increase the currently known sample of four Sednoids (Figure 2) and further refine this orbital boundary (Huang & Gladman 2024; Chen et al. 2025; Hu et al. 2025).

#### 3.2.5 Analysis summary

Overall, the results and discussion presented in Section 3 imply that an Earth-class outer planet with $a = 350$ au ($q = 200$ au; $i = 1°$ or 15°), or $a = 500$ au ($q = 200$ or 250 au; $i = 1$, 15, 30°), corresponding to PX2 models 7, 11, 12, 14, 15, 16, 17, and 18 in Table 2, can consistently produce detached, high-$i$, and extreme populations in excellent agreement with observational constraints. These results are consistent with the distant $a = 500$ au ($q \approx 200$ au, $i = 30°$) planets favoured in Lykawka & Ito (2023). Accounting for the constraint of low-$i$ Sednoids leads to a narrower inclination range for the putative planet, as $i = 30°$ becomes disfavoured.

In summary, our results strongly support the hypothesis that an undiscovered planet akin to Planet X2 exists in the outer realms of the Solar system.

### 3.3 Notes on the Planet 9 scenario

For completeness, we analysed the results of the systems generated under the Planet 9 scenario using the same criteria applied to the Planet X2 models discussed previously. Since the results for the inner trans-Neptunian region (< 50 au) were indistinguishable from those of the control, scattering-planet, and PX2 models, we focus our discussion here on the results for the distant region beyond 50 au. As expected, the perturbations of a super-Earth-mass Planet 9 substantially sculpted the orbital structure, generating conspicuous detached, high-$i$, and extreme populations (Figure 15). Although the production of these populations appears similar to some of our PX2 models, there are important differences. Specifically, because Planet 9 is significantly more massive and follows a less eccentric orbit, it produces unique orbital architectures beyond 50 au.

First, Planet 9 generated a substantial population of detached objects, resulting in the lowest scattered-to-detached population ratio (0.11) obtained across all our models. Second, the extreme population produced in the Planet 9 model was also exceptionally large, comprising a median of 82.4% of the entire population in the distant region (or 78.8% if considering the Sednoid population specifically). This represents the highest value among all models tested in this work.

Planet 9 also generated a distinct inclination distribution compared with the PX2 models, both in terms of concentrations in orbital space and high-$i$ population fractions. Notably, at the end of the simulations, the fraction of scattered objects with $i > 40°$ within the $a = 50$–200 au range reached a median of 4.3%. In comparison, PX2 models 14 and 17, which incorporated an outer planet with orbital parameters comparable to those of Planet 9, yielded fractions of 2.8% and 3.0%, respectively, while the control model yielded 2.7%. Scattered objects in this semimajor axis range are considered the primary source of ecliptic comets. Because the observed inclination distribution of ecliptic comets is concentrated at low inclinations ($i < 30$–$40°$), any excess of highly inclined precursors ($i > 40°$) in the source scattered disk would overproduce high-inclination comets, violating observational constraints (Volk & Malhotra 2008; Nesvorný et al. 2017; de Sousa et al. 2025). The Planet 9 configuration tested in this work (favoured by Brown & Batygin 2021) has an orbit significantly more compact than that considered by de Sousa et al. (2025). Those authors assumed a more distant and eccentric orbit with $a = 600$ au, $q = 420$ au, $i = 20°$, and $m = 7.5$ $M_{\oplus}$ to investigate the planet's potential stirring of the inclination distribution of ecliptic comets. Given our findings and these parameter differences, the compatibility of the Planet 9 scenario with the observed comet population warrants further investigation. Similarly, considering slightly different Planet 9 parameters ($a = 500$ au, $q = 375$ au, $i = 20°$, and $m = 5$ $M_{\oplus}$) that fall within the uncertainties determined by Brown & Batygin (2021), the results of Kaib et al. (2019) using the OSSOS observational sample suggest that Planet 9 generates an inclination distribution for the Centaur and scattering TNO populations ($q \lesssim 37$ au – both of which source ecliptic comets) that is more dynamically excited than observed.

Overall, the combination of a large mass and a relatively compact, moderately inclined orbit implies that Planet 9 produces dynamical signatures distinct from those of any Planet X2 model discussed previously. Given the potential overexcitation of the inclination distribution and the concerns outlined in Section 1, we defer a more detailed comparative study of the Planet 9 scenario to future work.

## 4. Observational constraints on the Planet X2 scenario from the Dark Energy Survey (DES)

Our Planet X2 models populate several orbital niches in the trans-Neptunian region consistent with intrinsic observational constraints (see Sections 1 and 3.2) and predict substantial populations of currently undiscovered TNOs. In this section, we evaluate whether simulated detections derived from these models are compatible with the observations from the Dark Energy Survey (DES). The DES dataset provides the benchmark characterised survey sample of distant TNOs available to date. Specifically, within the $a > 60$ au and $q > 25$ au region (matching the boundaries adopted in Section 4.1), the DES sample contains 136, 2, and 18 distant TNOs with $i > 0°$, $i > 45°$, and $q > 40$ au, respectively. However, while the DES dataset is well suited for our purposes, approximately 95% of this sample is confined to $a = 60$–$250$ au (Bernardinelli et al. 2022). This observational cut-off complicates efforts to constrain the vast distant populations generated in the Planet X2 scenario. Ultimately, the observed distributions at high inclinations, large perihelia, or $a > 200$ au remain limited by small-number statistics, restricting our ability to place tight constraints based strictly on comparisons between simulated detections and observations. We therefore treat these observational constraints as tentative complements to the intrinsic dynamical constraints discussed in previous sections.

To determine whether our models can be further differentiated, we used the DES Survey Simulator to directly compare our simulated populations with observations. The DES Survey Simulator is a computational tool developed by the DES collaboration to evaluate observational biases and survey completeness. It estimates the selection probability of TNOs based primarily on their orbital parameters and apparent magnitudes, while also incorporating secondary effects such as object colour. Additionally, the simulator accounts for detection biases that favour larger objects, objects near perihelion, and populations with inclinations that align with the DES footprint. For further details, we refer the reader to Bernardinelli et al. (2022, 2025).

We performed our analysis by comparing the orbital distributions of the simulator-detected model objects with those of the DES sample. As detailed in Section 2, the final populations obtained from successful systems were co-added for each specific model to provide a robust, phase-averaged underlying dataset. Next, to ensure a statistically sufficient number of model detections, we generated clones of the surviving objects for each model. We applied random variations to the original orbital elements (±10%, ±1%, and ±1% for $a$, $i$, and $q$, respectively) and uniformly randomised their orbital angles (Ω, ω, and M) from 0° to 360° to generate 60 million synthetic particles per model as simulator input. These cloning and co-addition procedures are consistent with previous studies (e.g., Pike et al. 2017; Kaib et al. 2019; Lykawka & Ito 2023; Kaib et al. 2024). This approach typically yielded several thousand detections per model to compare with the distant TNOs found by DES. For our synthetic populations, we employed the nominal absolute magnitude ($H$) distribution statistically validated by Bernardinelli et al. (2025) across all major TNO populations constrained by DES observations. This model consists of two rolling power-law $H$ distributions, corresponding to the two colour families found in DES data (near-IR bright, NIRB, and near-IR faint, NIRF), where the NIRF component is redder and corresponds to the colour group occupied by the cold classicals. The two populations are then mixed following the corresponding fraction of each color group, adopted nominally as 70% NIRB, matching the hot classical, scattering and detached populations from Bernardinelli et al. (2025). As illustrated in Figure 16, our adopted $H$-distribution differs from the dual-slope distributions constrained by earlier OSSOS studies (specifically the 'divot' and 'knee' distributions; see Shankman et al. 2013; Shankman et al. 2016; Lawler et al. 2018b), but is similar to the $H$-distribution derived from OSSOS for the hot classical belt (Petit et al. 2023; this similarity supports the idea that the scattered and detached populations share a common dynamical origin with the hot classicals). Our choice is also consistent with the fact that the majority of the intrinsic distant TNOs and Planet X2 model objects belong to the scattered and detached populations. Consequently, this $H$-distribution is self-consistent and appropriate for analysing the distant trans-Neptunian populations in this work.

Finally, to evaluate the model detections with the DES sample, we employ both Kolmogorov-Smirnov (KS) and Anderson-Darling (AD) statistical tests at a 5% significance level, requiring the $p$-values from both tests to fall below 5% to reject the null hypothesis that the simulated distribution of semimajor axis, inclination, or perihelion is drawn from the same parent population as the DES observational sample. For brevity, we report the $p$-values from both tests as a range hereafter. To quantify the sampling uncertainties arising from the small DES sample size, our analysis also incorporates 95% confidence envelopes around the survey-biased model distributions. These were derived via Monte Carlo resampling, generating 1000 mock surveys matching the DES sample size (drawn without replacement) to determine the expected sampling variance.

### 4.1 Comparing simulation results with observations

To determine whether the simulated detection distributions from our preferred Planet X2 models (Section 3.2.5) are statistically compatible with DES observations, we considered all objects with orbits falling within two specific regions of interest. The first region was set at $a$ = 60–100 au, an area dynamically shaped by Neptune's gravitational and resonant perturbations during and after planetary migration. Thus, the orbital structure in this region serves as a proxy to validate the Neptune grainy-migration model. The second region was defined within $a$ = 60–250 au (where adopting alternative outer boundaries at 200 or 300 au yielded negligible differences, given the small number of TNOs in the DES sample beyond 200 au). This region partially overlaps the domain perturbed by Planet X2, allowing us to tentatively constrain the generated populations. Although we initially applied a $q$ = 25–55 au filter to both regions to match the DES dataset, our analysis and discussion below focus specifically on the more stable populations with $q$ = 35–55 au (consistent with the approach adopted in Nesvorny et al. 2023). The region at $a < 50$ au was omitted from this analysis, as modelling it requires a specifically tuned survey simulator to accurately distinguish between the cold and hot classical components and the inner resonant populations. We also excluded the 5:2 MMR population to avoid complications associated with modelling its population fraction relative to inner resonances at $a < 50$ au. Consequently, our inner boundary was fixed at 60 au.

Within the compact $a$ = 60–100 au region, the semimajor axis, inclination, and perihelion distributions of detected objects across all tested models are closely similar. As anticipated, this result reinforces the premise that Neptune's perturbations dominate the dynamical sculpting of the inner trans-Neptunian region. These distributions are also broadly consistent with the DES observations, yielding $p$-values above the 5% significance level[6]. Because this orbital structure is primarily driven by Neptune, our findings suggest that the grainy migration model, as modelled here and in representative studies such as Kaib et al. (2024), satisfactorily explains the bulk orbital architecture within 60–100 au. Figure 17 illustrates a representative comparison between the control model and Planet X2 model 15.

The comparison of model detections with DES observations across the extended $a$ = 60–250 au region strongly constrains our models. The cumulative $a$, $i$, and $q$ distributions of simulated detections for our best models are illustrated in Figures 18 and 19. First, we found that the semimajor axis distribution of all models is statistically incompatible with the DES sample, yielding $p$-values below 0.1%, well below our 5% rejection threshold. Because the models predict relatively fewer detections at larger semimajor axes, the simulated distribution systematically deviates from the observed sample. This highlights the long-standing semimajor axis mismatch identified in previous studies comparing distant populations with observations (Lawler et al. 2017, 2018b; Beaudoin et al. 2023; Lykawka & Ito 2023; Nesvorný et al. 2023). Nevertheless, a potential resolution involves introducing a primordial scattered disk that initially extended beyond 50 au. This configuration aligns with models predicting the existence of such a structure after giant-planet formation (prior to the instability) and during the onset of the instability (de Sousa et al. 2020; Huang et al. 2022b; Hu et al. 2025). Moreover, our earlier work (Lykawka & Ito 2023) and DES Survey Simulator tests performed here indicate that the inclination and perihelion distributions are primarily governed by perturbations from Neptune and Planet X2, with the influence of the latter also depending on its mass and orbital parameters. Consequently, an extended initial semimajor axis distribution should better populate distant reservoirs without significantly altering the $i$ and $q$ profiles. Furthermore, while our Planet X2 models yield 1D intrinsic orbital distributions comparable to the 'Cluster2' model of Nesvorný et al. (2023, in which a single, exceptionally close stellar flyby dynamically shapes the distant trans-Neptunian region), our results do not reproduce their simulated semimajor axis detections in the DES Survey Simulator. This discrepancy likely stems from the adopted $H$-distribution (governing heliocentric detection limits), multidimensional orbital correlations ($a$–$q$–$i$) in the detached region ($q > 40$ au) driven by an impulsive stellar kick versus secular Planet X2 perturbations, particle-sampling stochasticity in their single N-body realisation, or a combination of these factors. Disentangling these effects across dynamical sculpting scenarios remains an open problem for future work.

[6] Only PX2 model 8 exhibited incompatibility in semimajor axis ($p$-values ~2%), whereas PX2 model 18 exhibited incompatibility in inclination ($p$-values ~0–3%).

Regarding the other orbital elements, while the control model underpredicts the number of objects with $i < 20°$, it remains statistically compatible with observations ($p$-values ~8–24%). With a few exceptions (discussed below), the PX2 models typically produced detected inclination distributions compatible with the DES sample. However, severe observational biases at high inclinations restricted the sample size of these highly inclined populations, preventing meaningful statistical comparisons at the high-inclination tail. Finally, the perihelion distributions across all tested models, including the control, exhibited excellent agreement with observations.

Overall, models incorporating a Planet X2 with $q$ = 200 au and $i$ = 1° or 15° yielded better agreement with the observed inclination distributions. At $a$ = 500 au, configurations with $q$ = 200 au systematically outperformed those with $q$ = 250 au ($p$-values of ~4–46% vs ~0–18%). Furthermore, across all ($a$, $q$) combinations, an inclination of $i$ = 30° consistently yielded the poorest statistical fits, reinforcing our earlier finding that such high planetary inclinations are disfavoured because they also overexcite and deplete the low-inclination detached reservoir (Section 3.2).

We conclude that the Planet X2 scenario satisfies several intrinsic trans-Neptunian constraints and successfully reproduces the DES perihelion and inclination distributions across $a$ = 60–250 au. Although fully reconciling the detected semimajor axis distribution with the DES detections requires further model refinement, overall, these results provide strong support for the presence of an Earth-class planet in the distant Solar system.

## 5. Summary

We performed extensive N-body simulations to investigate the 4.5 Gyr orbital evolution of the primordial trans-Neptunian region. Our baseline model incorporated the four giant planets, a planetesimal disk, and 1500 Pluto-class bodies. Using this framework, we tested and compared two distinct mechanisms for sculpting the outer Solar system: the influence of primordial scattering planets (Mars- to Earth-mass embryos initially embedded within 30 au) and the presence of an undiscovered, Earth-class outer planet with a mass comparable to Earth—designated as the Planet X2 scenario (alongside a reference super-Earth-mass Planet 9 model).

Among the tested hypotheses, our results demonstrate that the Planet X2 scenario best reproduces the observed orbital properties of the trans-Neptunian region. The alternative scenario involving primordial scattering planets proved highly inefficient, as these bodies are generally short-lived (with dynamical lifetimes of less than a few tens of Myr) and fail to adequately populate distant reservoirs, such as the detached and extreme populations. Our detailed analysis revealed that while Neptune's migration is essential for establishing the basic orbital structure within ~50–100 au, the long-term gravitational influence of Planet X2 is necessary to reproduce the orbital architecture observed beyond 50 au. The best-performing Planet X2 models effectively generate detached objects with perihelia $q > 40$ au and high-$i$ populations with $i > 45°$. Furthermore, these models are particularly efficient at producing extreme populations with elevated perihelia or inclinations (while preserving an $i < 30°$ component required for the Sednoid regime at $a$ = 200–1500 au and $q > 60$ au), thereby populating the observed orbital $a$–$q$–$i$ niches that remain unexplained by published migration models, as confirmed by our control model. Consequently, our findings indicate that the Planet X2 scenario can account for all primary dynamical populations in the trans-Neptunian region, including the classical, resonant, scattered, detached, and high-$i$ populations, along with their extreme members. Specifically, the model replicates the orbits of all known extreme members, such as (90377) Sedna, 2012 $VP_{113}$, and 2015 $BP_{519}$, as well as those located at semimajor axes $a > 500$ au. When compared with the sample data from the Dark Energy Survey (DES), Planet X2 models yielded detection outcomes that are potentially compatible with DES observations. However, the shortage of model detections at larger semimajor axes within the studied $a$ = 60–250 au region suggests further model refinement is required in the future (e.g., considering the existence of primordial extended scattered disks).

Our results support the existence of an undiscovered outer planet on a distant, stable orbit. By analysing the impact of various Planet X2 parameters, we established $a$ = 350–500 au as the optimal configuration to simultaneously satisfy major observational constraints. Furthermore, within the parameter space tested in this work, the best models tentatively constrain the planet's perihelion to 200 au (rather than 250 au), its inclination to 1–15° (rather than 30°), and its mass to 2 $M_\oplus$ (rather than 1 $M_\oplus$). While simulating the origin of Planet X2 is beyond the scope of our N-body integrations, several mechanisms provide potential pathways for the formation of an Earth-class planet on a distant, high-perihelion orbit. Plausible scenarios to explore in future work include in situ growth (Kenyon & Bromley 2015, 2016), outward gravitational scattering followed by eccentricity damping via gas (Bromley & Kenyon 2014) or planetesimal dynamical friction (Eriksson et al. 2018; Bailey & Fabrycky 2019), alternative embryo mass spectra for multiple scattering (Silsbee & Tremaine 2018; this work), and early distant stellar flybys (Izidoro et al. 2025).

The Planet X2 scenario also predicts specific orbital signatures that provide observationally testable markers for ongoing and upcoming surveys, notably the Legacy Survey of Space and Time (LSST) at the Vera C. Rubin Observatory (Ivezić et al. 2019; Kurlander et al. 2025). Such wide-field surveys are well placed either to directly detect the putative planet or to test the statistical significance of the orbital features predicted for the detached, high-$i$, and extreme populations beyond 100 au.

## Acknowledgments

We thank the anonymous referee for detailed, insightful, and constructive comments, which significantly improved the quality and clarity of this paper. The numerical simulations presented here were carried out in part on the general-purpose PC cluster at the Center for Computational Astrophysics in the National Astronomical Observatory of Japan. We are grateful for the computing time allocated to this project. This work was supported by JSPS KAKENHI Grant Number JP23K03482. P. H. B. acknowledges support from the DIRAC Institute in the Department of Astronomy at the University of Washington. The DIRAC Institute is supported through generous gifts from the Charles and Lisa Simonyi Fund for Arts and Sciences, and the Washington Research Foundation.

## Data Availability

The data presented in this study substantiate the findings reported herein. Observational data can be accessed through the AstDyS database at https://newton.spacedys.com/astdys/. The full DES data catalog, the DES Survey Simulator, and the color/magnitude model are available on GitHub at https://github.com/bernardinelli/des_tno_catalog, https://github.com/bernardinelli/DESTNOSIM, and https://github.com/bernardinelli/des_tno_likelihood, respectively. Supplementary data and simulation outputs are available from the corresponding author upon reasonable request.

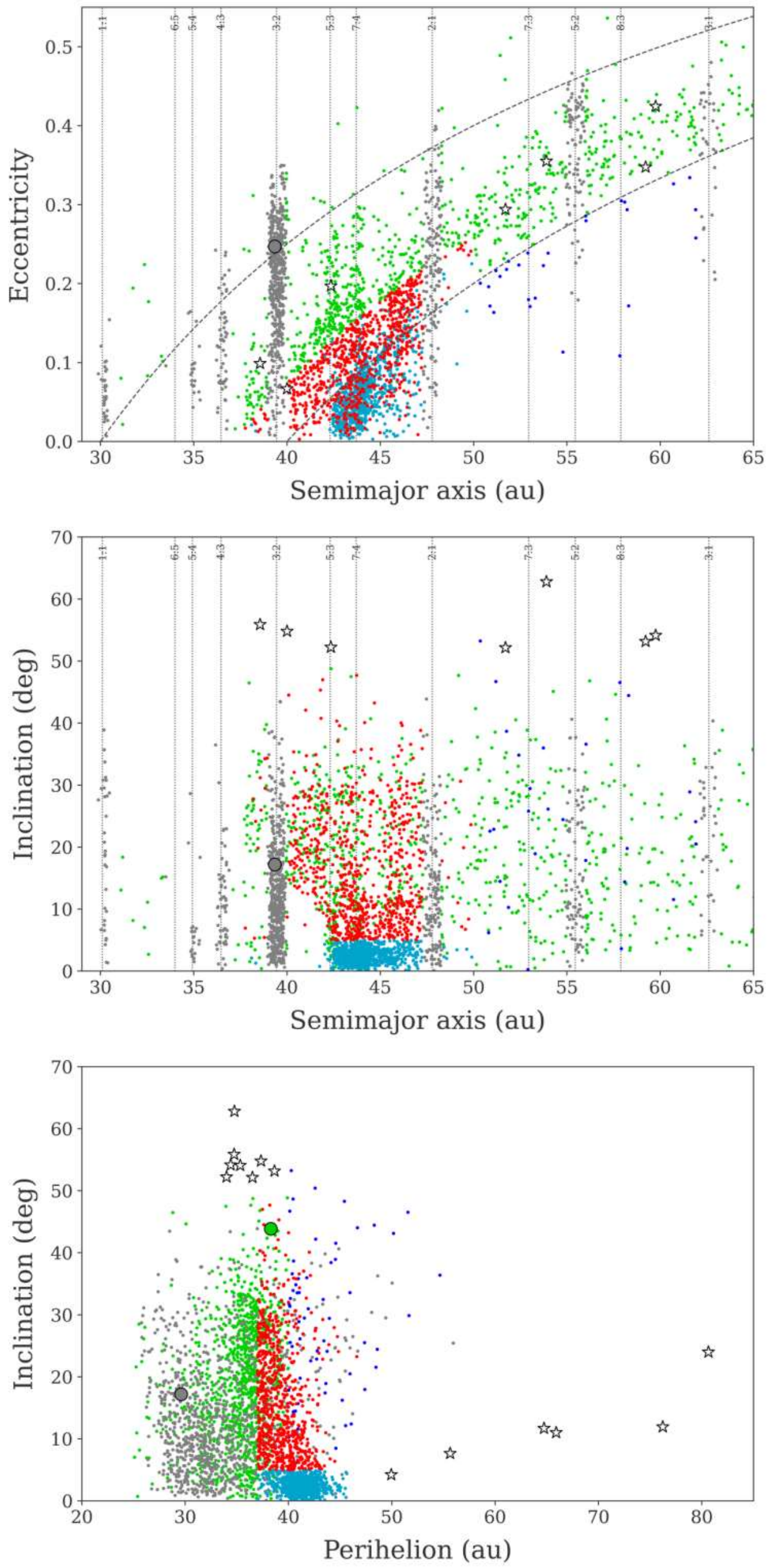

**Figure 1.** Orbital distribution of trans-Neptunian objects (TNOs) with small orbital uncertainties obtained from the AstDyS database (the nominal orbits' 1σ uncertainties for the semimajor axis and perihelion are below 5%, with medians of 0.06% and 0.007%, respectively). TNOs are dynamically classified as potentially resonant (grey), cold classical (dark cyan), hot classical (red), scattered (green), and detached (blue). Resonance status is determined by objects exhibiting libration in a mean motion resonance with Neptune for >50% of the 10 Myr-integration timescale. Classical TNOs concentrate at semimajor axes $a \approx 40$–50 au and perihelion distances $q > 37$ au. Scattered and detached TNOs are non-classical objects with $q \lesssim 40$ au and $q \gtrsim 40$ au, respectively. The cold/hot classical and scattered/detached dynamical boundaries are approximations adopted for straightforward categorisation. Extreme TNOs are indicated by stars. Pluto and Eris are represented by large circles and colour-coded according to the same classification. See Sections 1 and 2 for further details.

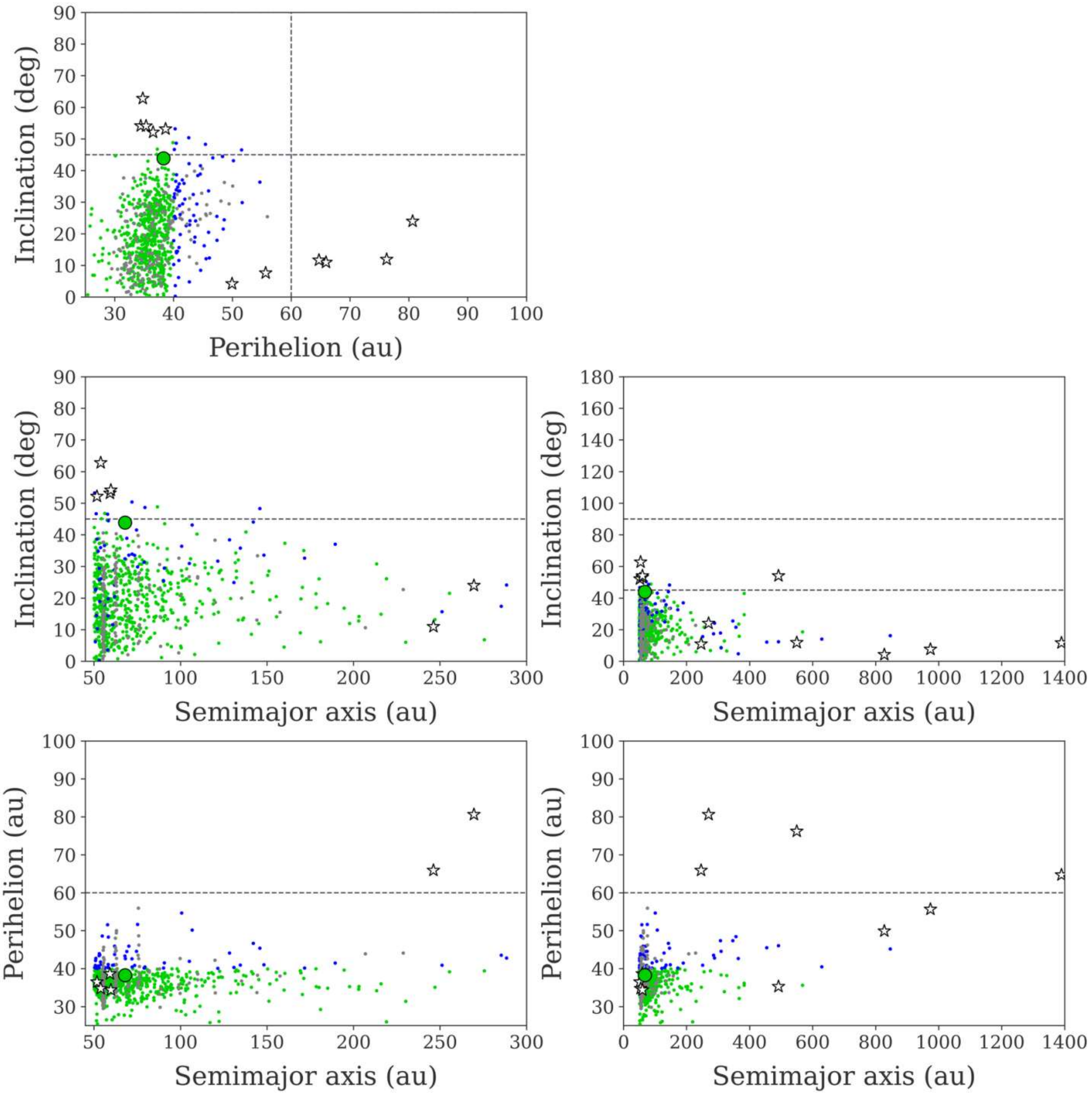

**Figure 2.** Same as Figure 1, but showing only distant TNOs with $a > 50$ au and $q > 25$ au. Dashed lines mark key dynamical boundaries. The condition $q > 60$ au indicates highly detached orbits (characteristic of Sednoids at $a > 200$ au), whereas $i > 45°$ identifies high-inclination TNOs. Additionally, $i = 90°$ separates prograde and retrograde orbits. See the main text for other details.

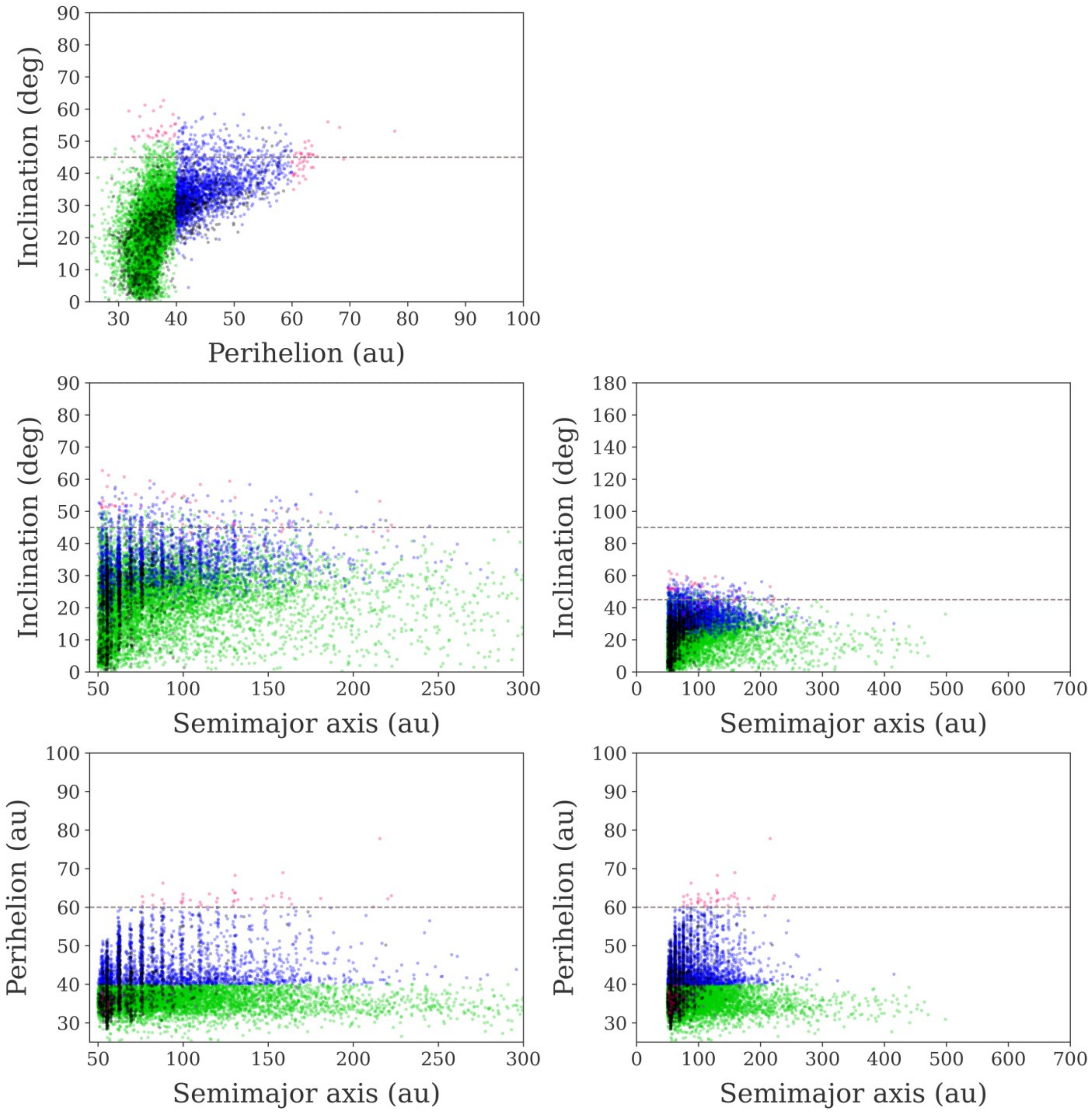

**Figure 3.** Orbital distribution of distant simulated objects ($a$ > 50 au, $q$ > 25 au) surviving 4.5 Gyr of dynamical evolution across the 50 representative systems obtained in our control grainy migration model (without additional planets). Objects are colour-coded by dynamical class: resonant (black), scattered (green), and detached (blue), with extreme members shown in violet (see Sections 1 and 2 for classification details). To facilitate direct comparison across all modelled scenarios in this work, the panel axis ranges and symbol styles are kept consistent with those in Figures 4, 5 and 6–15; quantitative comparisons between the orbital distributions shown here and those in subsequent figures are detailed in Section 3.

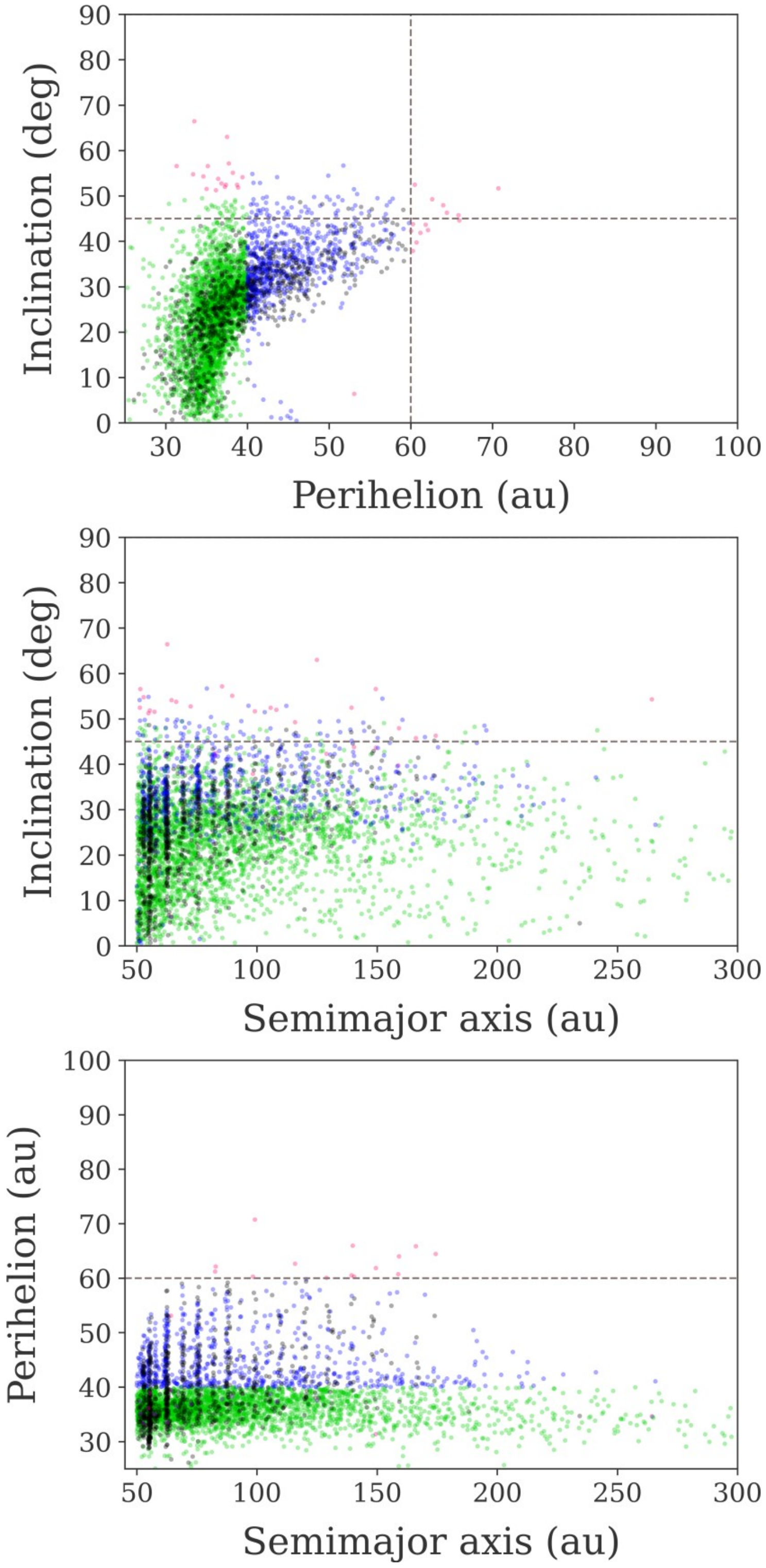

**Figure 4.** Similar to Figure 3, but showing the distant objects that survived 4.5 Gyr of dynamical evolution across the 25 successful systems obtained from the Emb-1M3 model, in which the scattering planets' lifetimes were less than 30 Myr. This model incorporates grainy migration alongside a single 3xMars-mass scattering planet (Section 3.1 and Table 2).

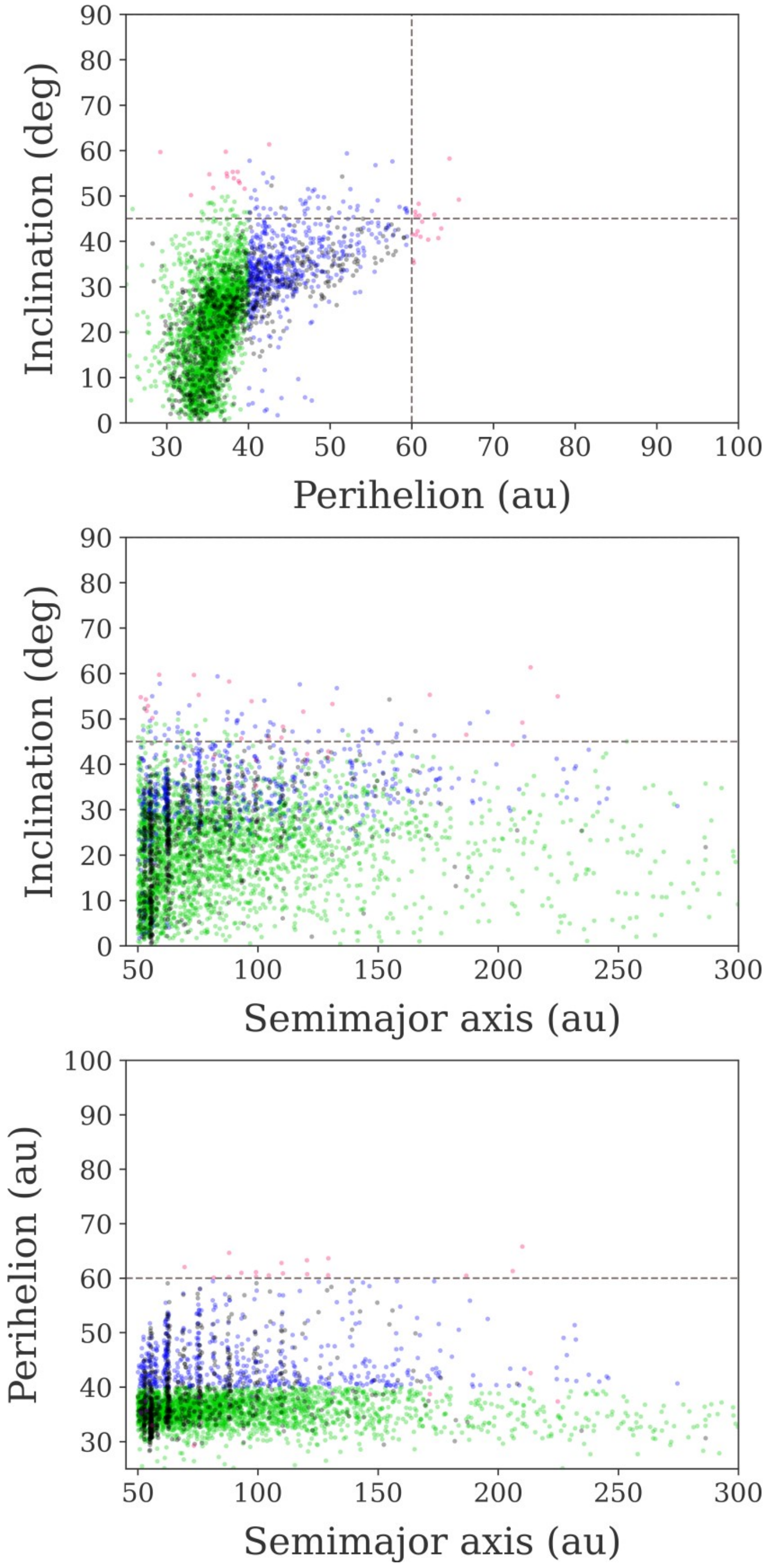

**Figure 5.** Similar to Figure 3, but showing the distant objects that survived 4.5 Gyr of dynamical evolution across the 17 successful systems obtained from the Emb-1E model, in which the scattering planets' lifetimes were less than 30 Myr. This model incorporates grainy migration alongside a single Earth-mass scattering planet (Section 3.1 and Table 2).

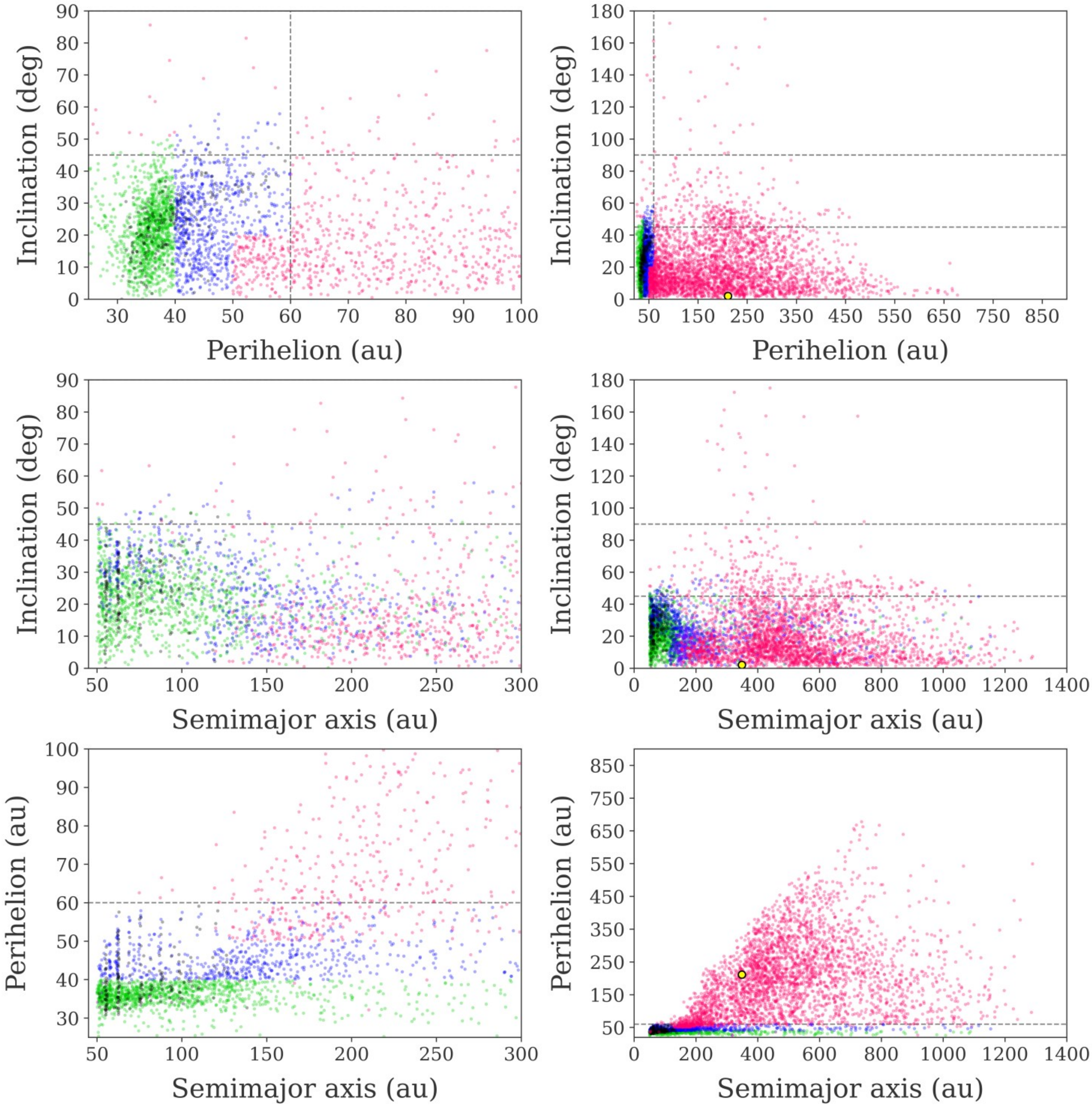

**Figure 6.** Orbital distribution of distant objects ($a > 50$ au, $q > 25$ au) surviving 4.5 Gyr of dynamical evolution across representative systems obtained from Planet X2 model 7. This model incorporates grainy migration alongside a single Earth-class outer planet with $a = 350$ au, $i = 1°$, $q = 200$ au, and $m = 2$ $M_\oplus$ (Section 3.2 and Table 2). The large yellow dot represents the orbit of Planet X2. The symbols and colour-coding for the remaining populations are consistent with those described in Figure 3.

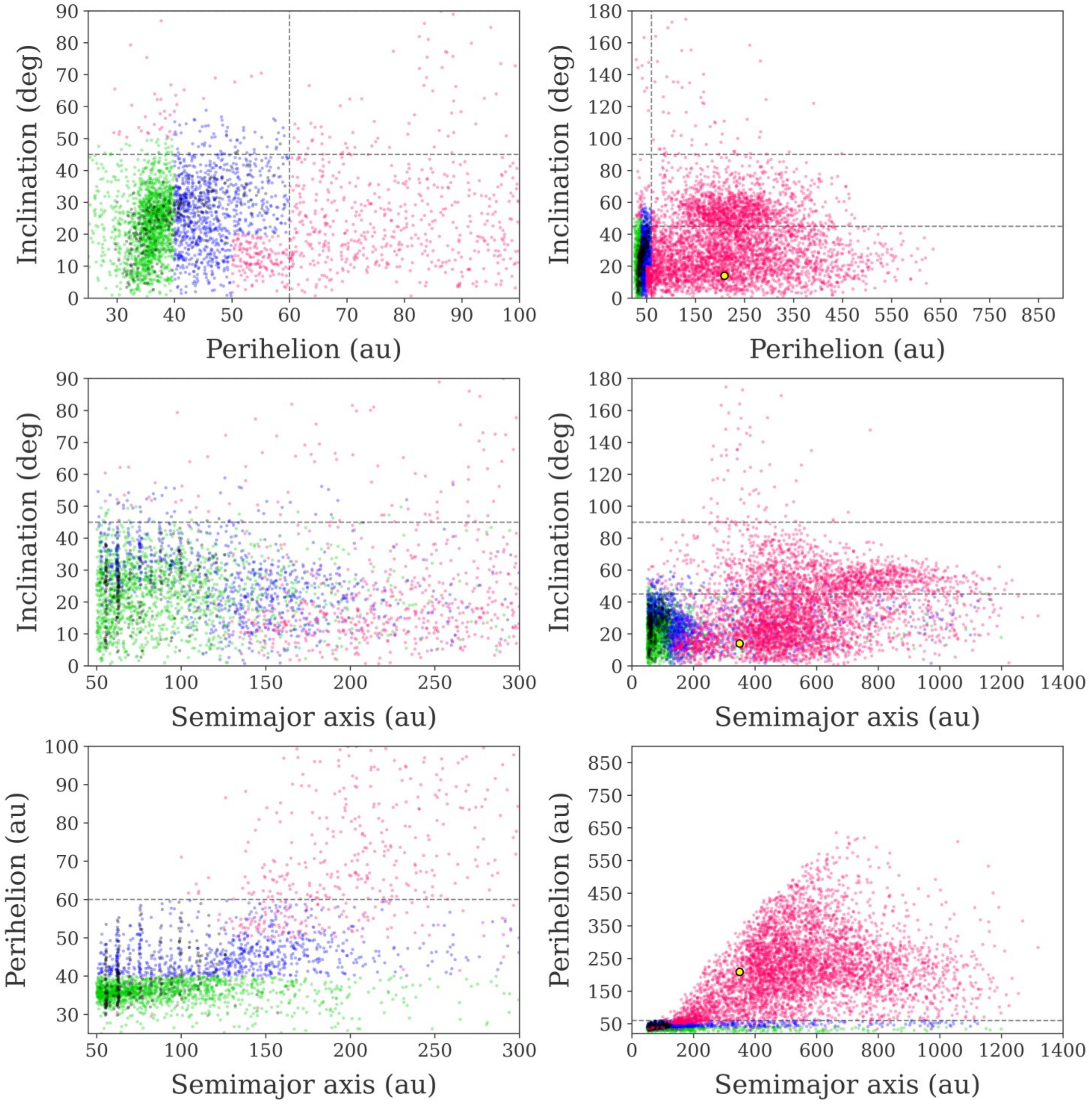

**Figure 7.** Same as Figure 6, but showing the results obtained from PX2 model 14. In this model, Planet X2 has $a$ = 350 au, $i$ = 15°, $q$ = 200 au, and $m$ = 2 $M_\oplus$.

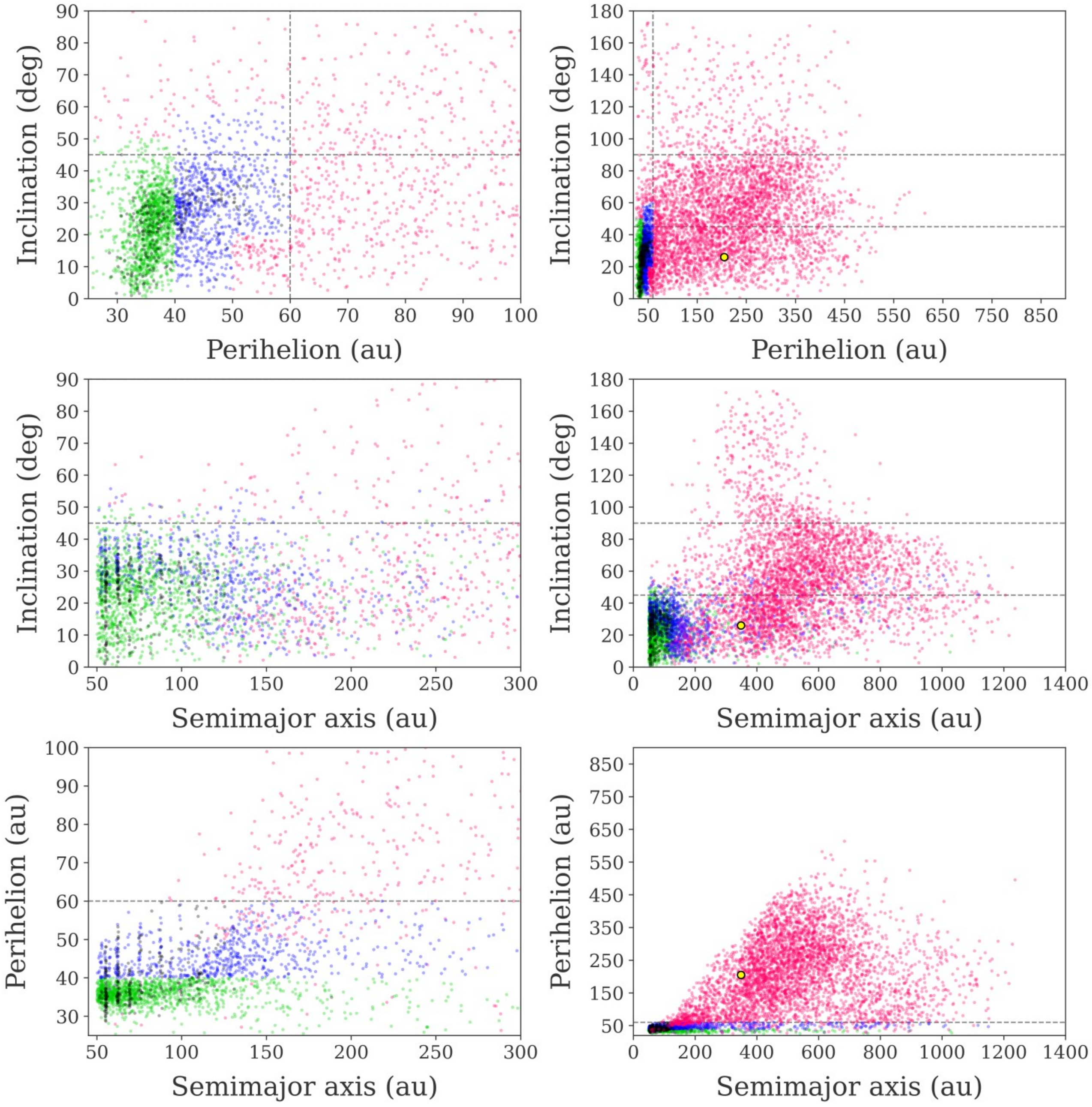

**Figure 8.** Same as Figure 6, but showing the results obtained from PX2 model 8. In this model, Planet X2 has $a$ = 350 au, $i$ = 30°, $q$ = 200 au, and $m$ = 2 $M_\oplus$.

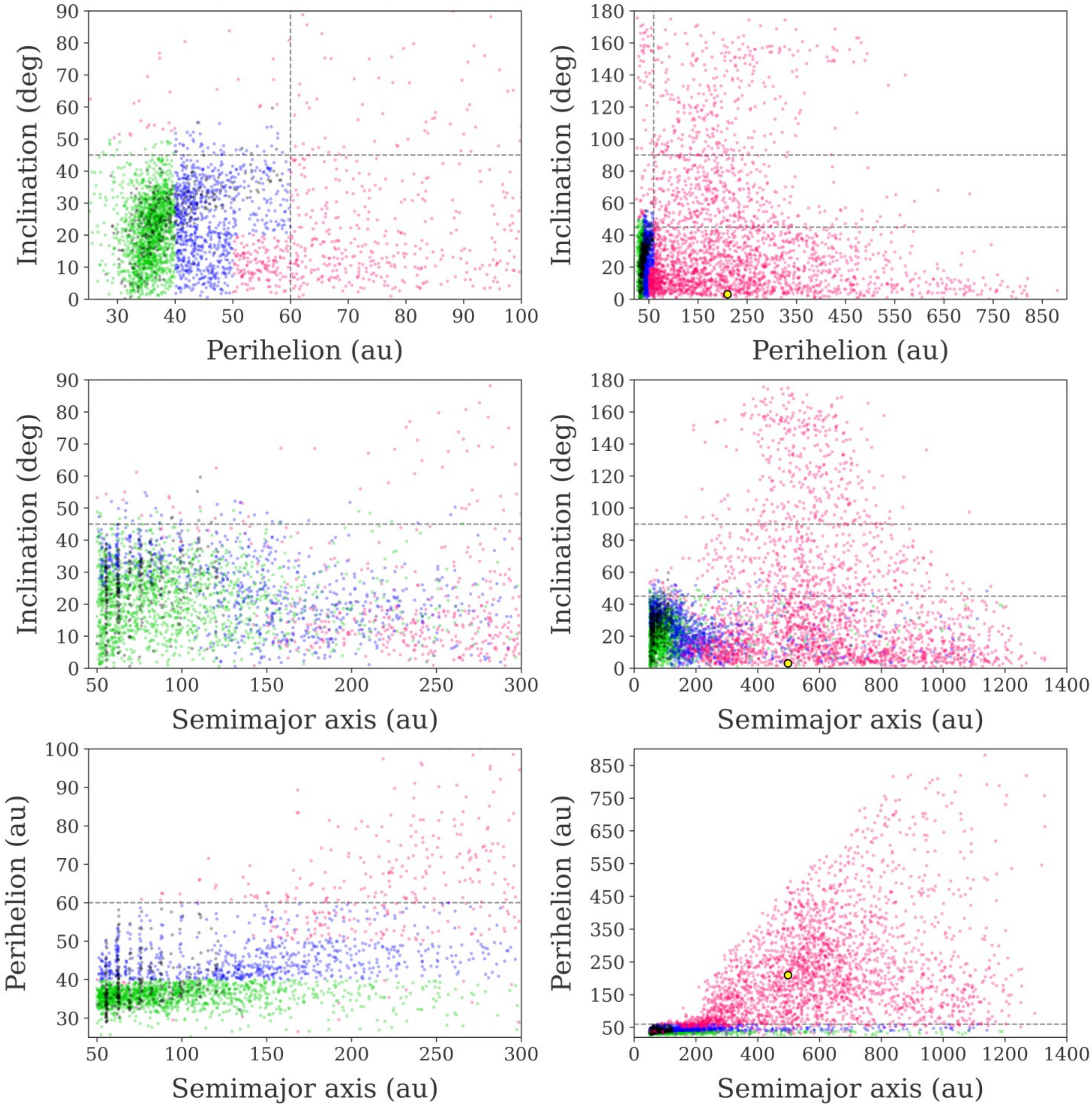

**Figure 9.** Same as Figure 6, but showing the results obtained from PX2 model 11. In this model, Planet X2 has $a$ = 500 au, $i$ = 1°, $q$ = 200 au, and $m$ = 2 $M_{\oplus}$.

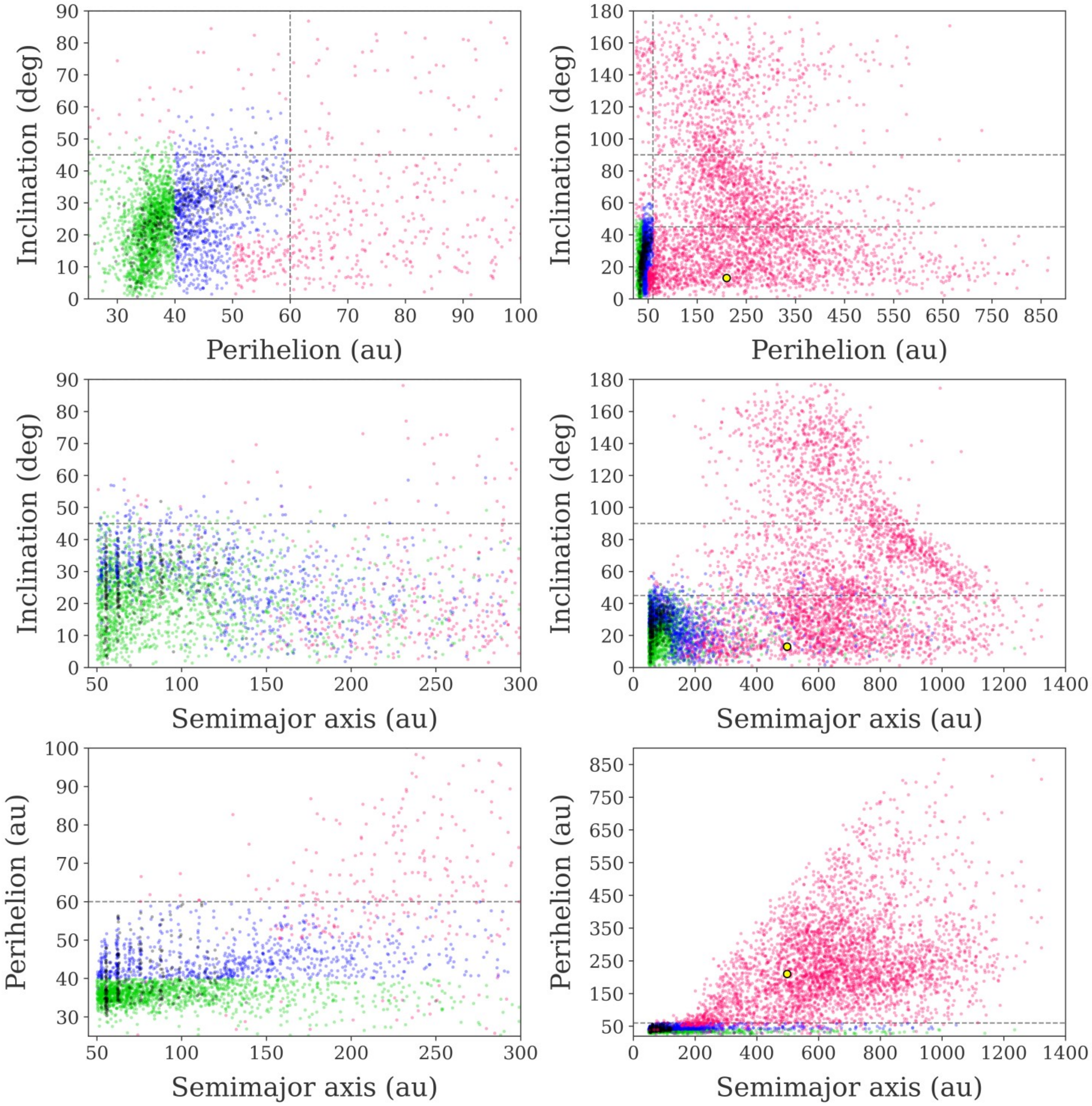

**Figure 10.** Same as Figure 6, but showing the results obtained from PX2 model 15. In this model, Planet X2 has $a$ = 500 au, $i$ = 15°, $q$ = 200 au, and $m$ = 2 $M_{\oplus}$.

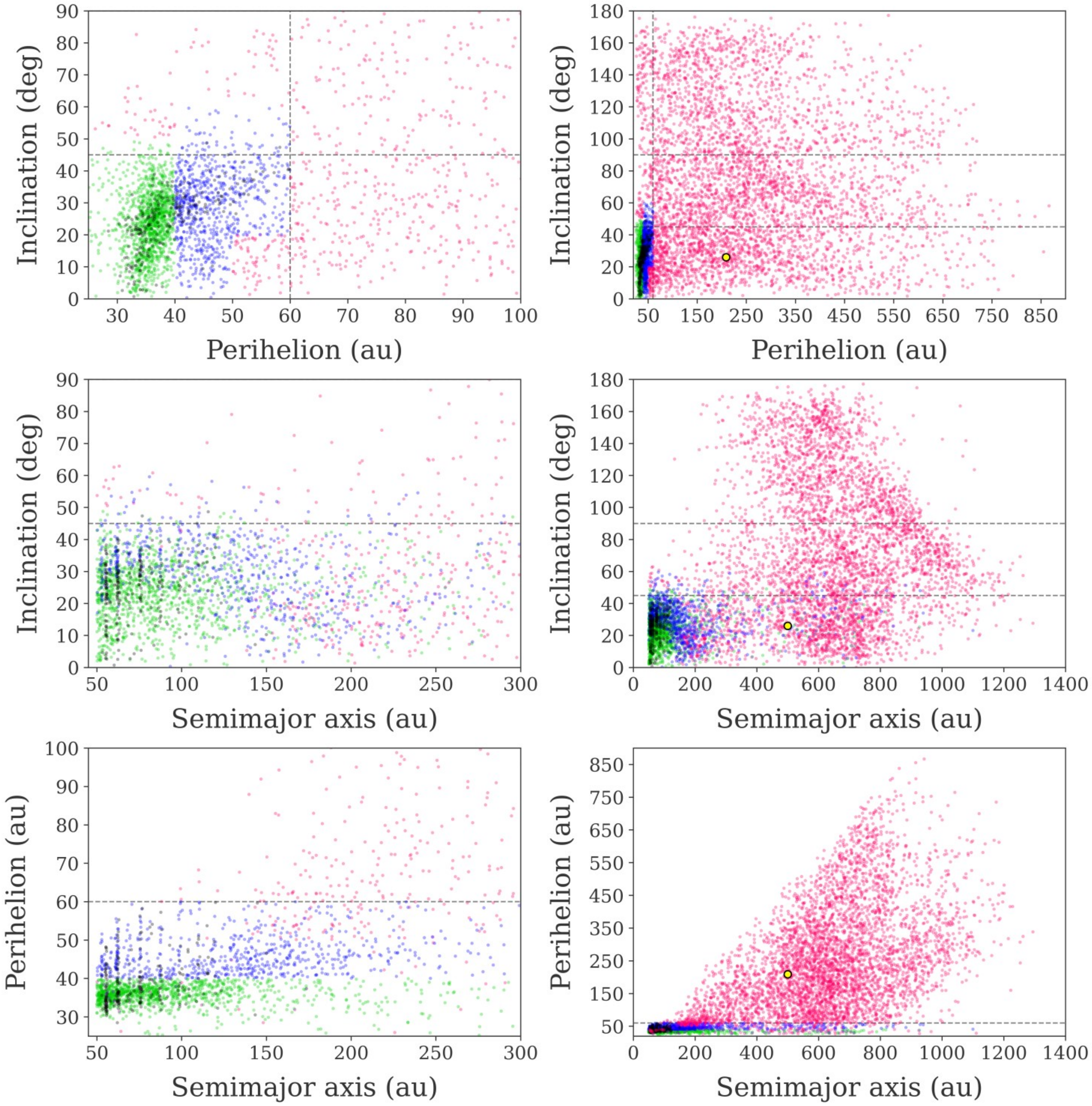

**Figure 11.** Same as Figure 6, but showing the results obtained from PX2 model 12. In this model, Planet X2 has $a$ = 500 au, $i$ = 30°, $q$ = 200 au, and $m$ = 2 $M_\oplus$.

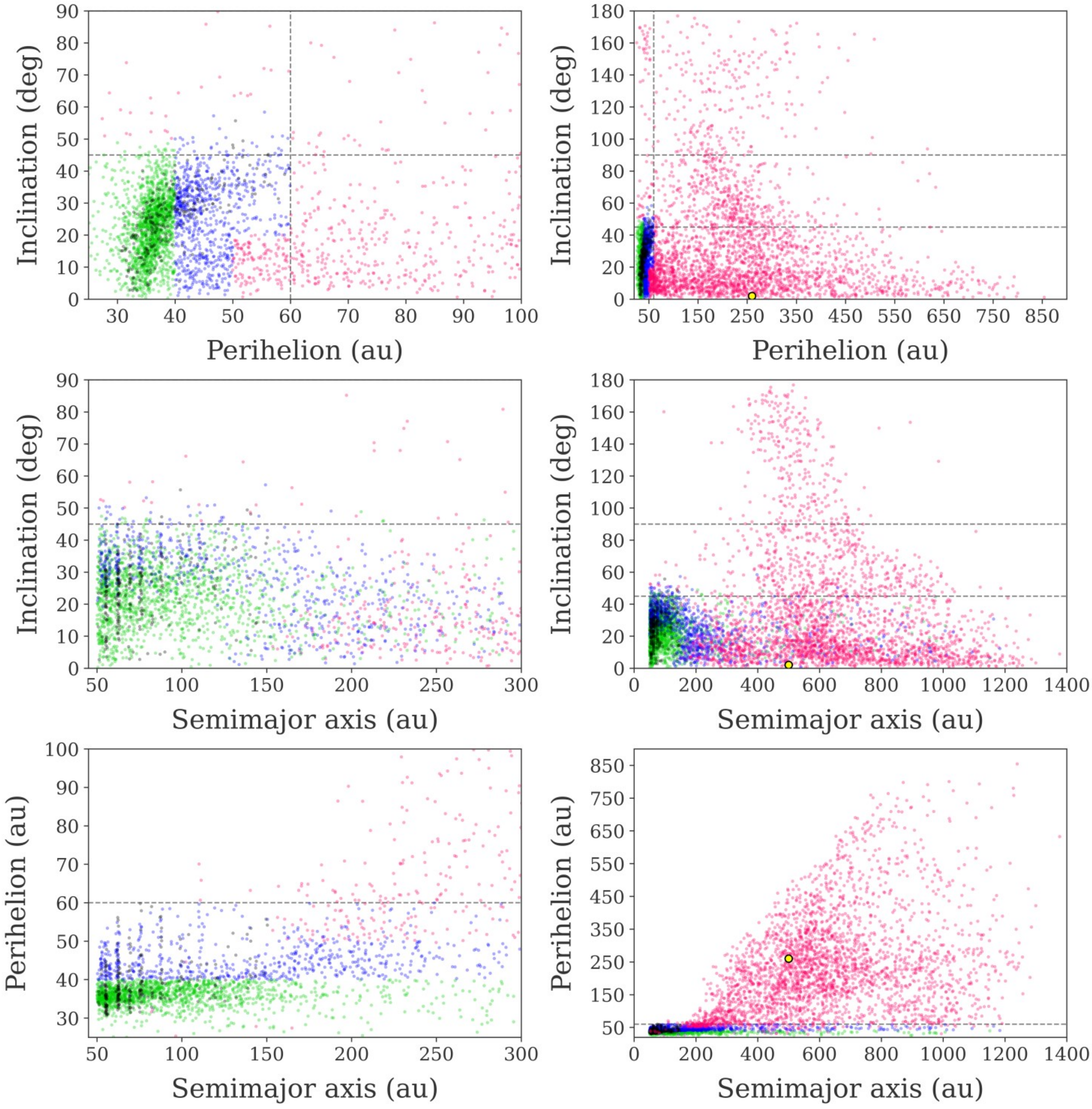

**Figure 12.** Same as Figure 6, but showing the results obtained from PX2 model 16. In this model, Planet X2 has $a$ = 500 au, $i$ = 1°, $q$ = 250 au, and $m$ = 2 $M_\oplus$.

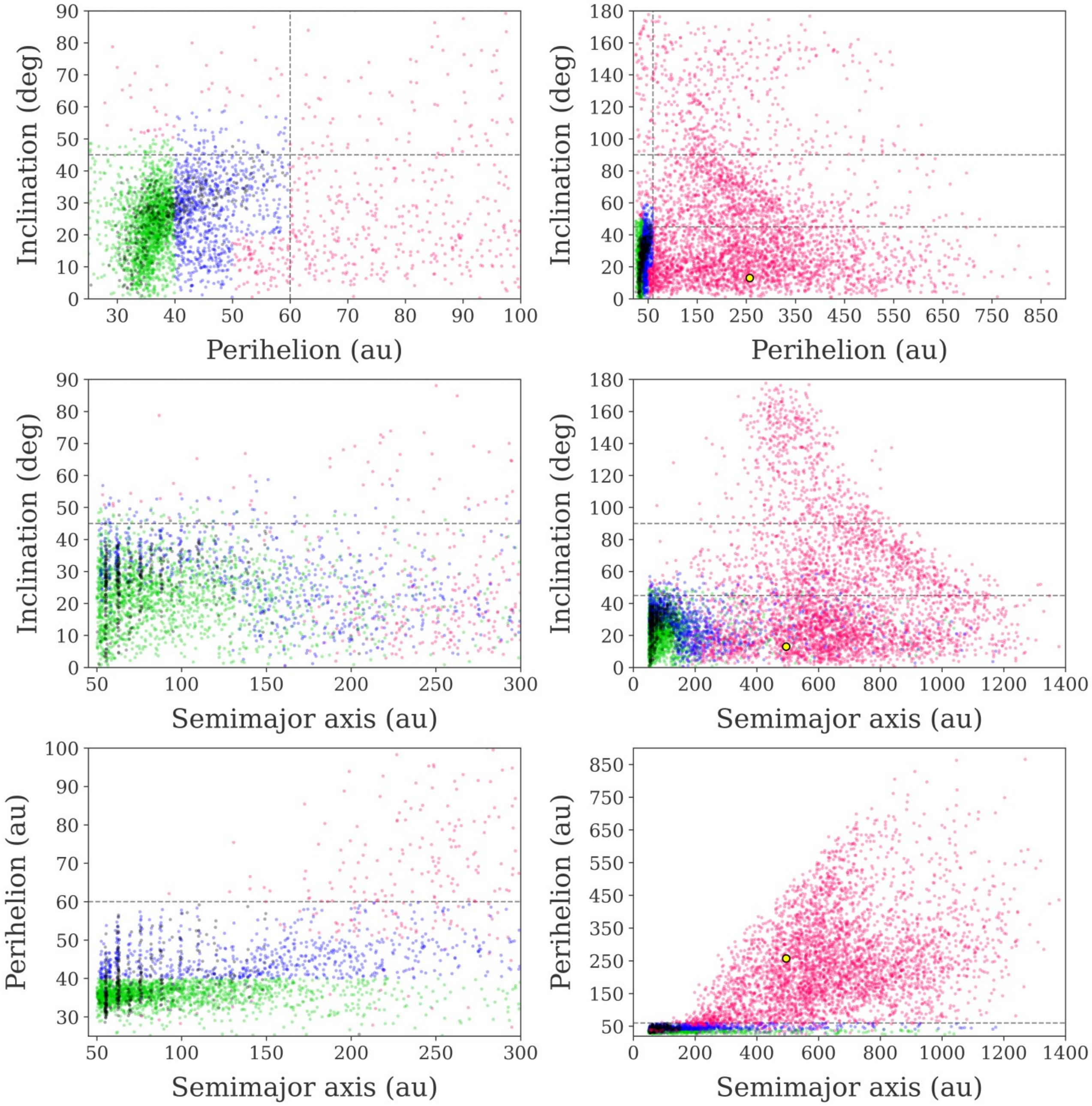

**Figure 13.** Same as Figure 6, but showing the results obtained from PX2 model 17. In this model, Planet X2 has $a$ = 500 au, $i$ = 15°, $q$ = 250 au, and $m$ = 2 $M_{\oplus}$.

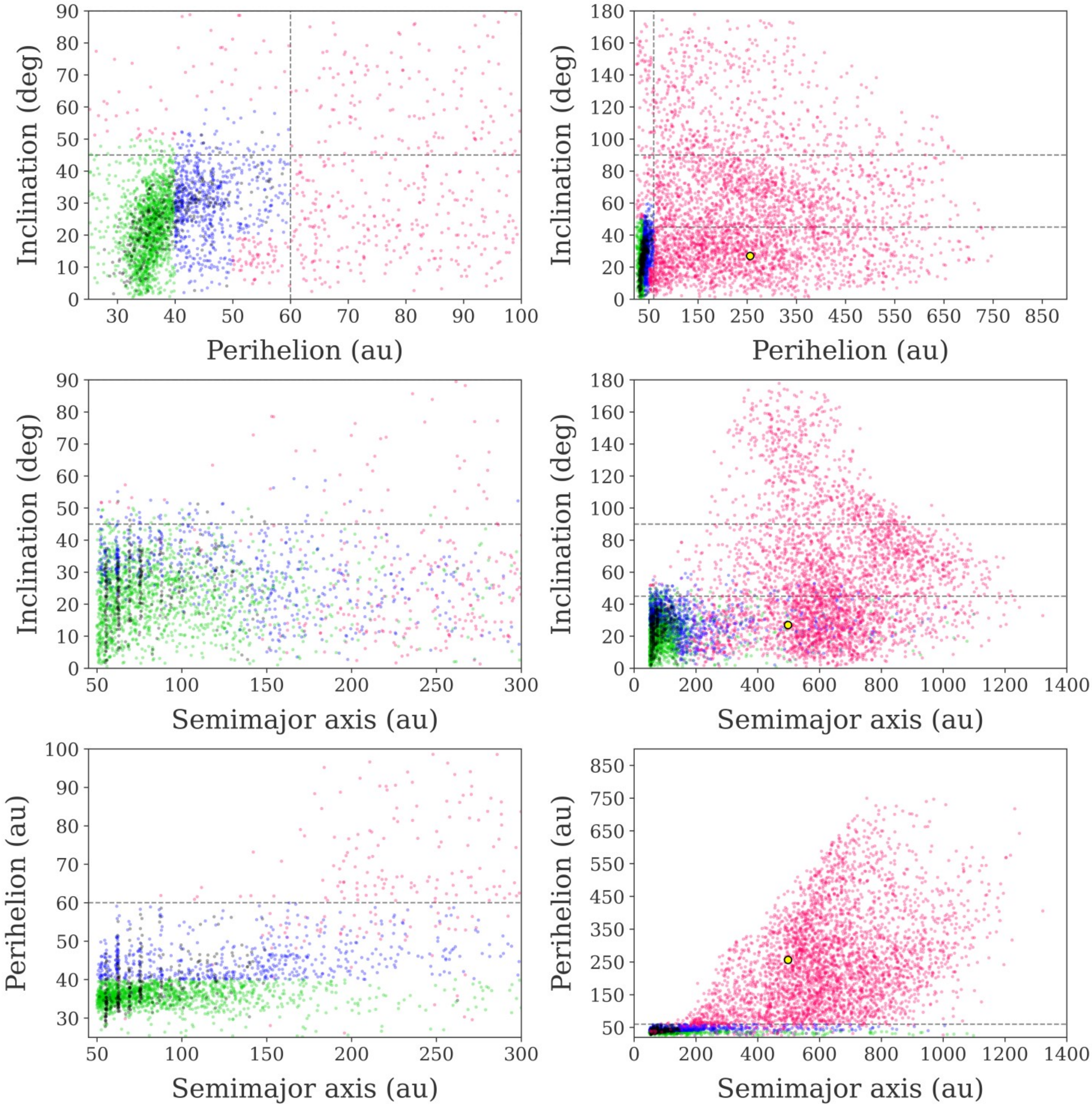

**Figure 14.** Same as Figure 6, but showing the results obtained from PX2 model 18. In this model, Planet X2 has $a$ = 500 au, $i$ = 30°, $q$ = 250 au, and $m$ = 2 $M_\oplus$.

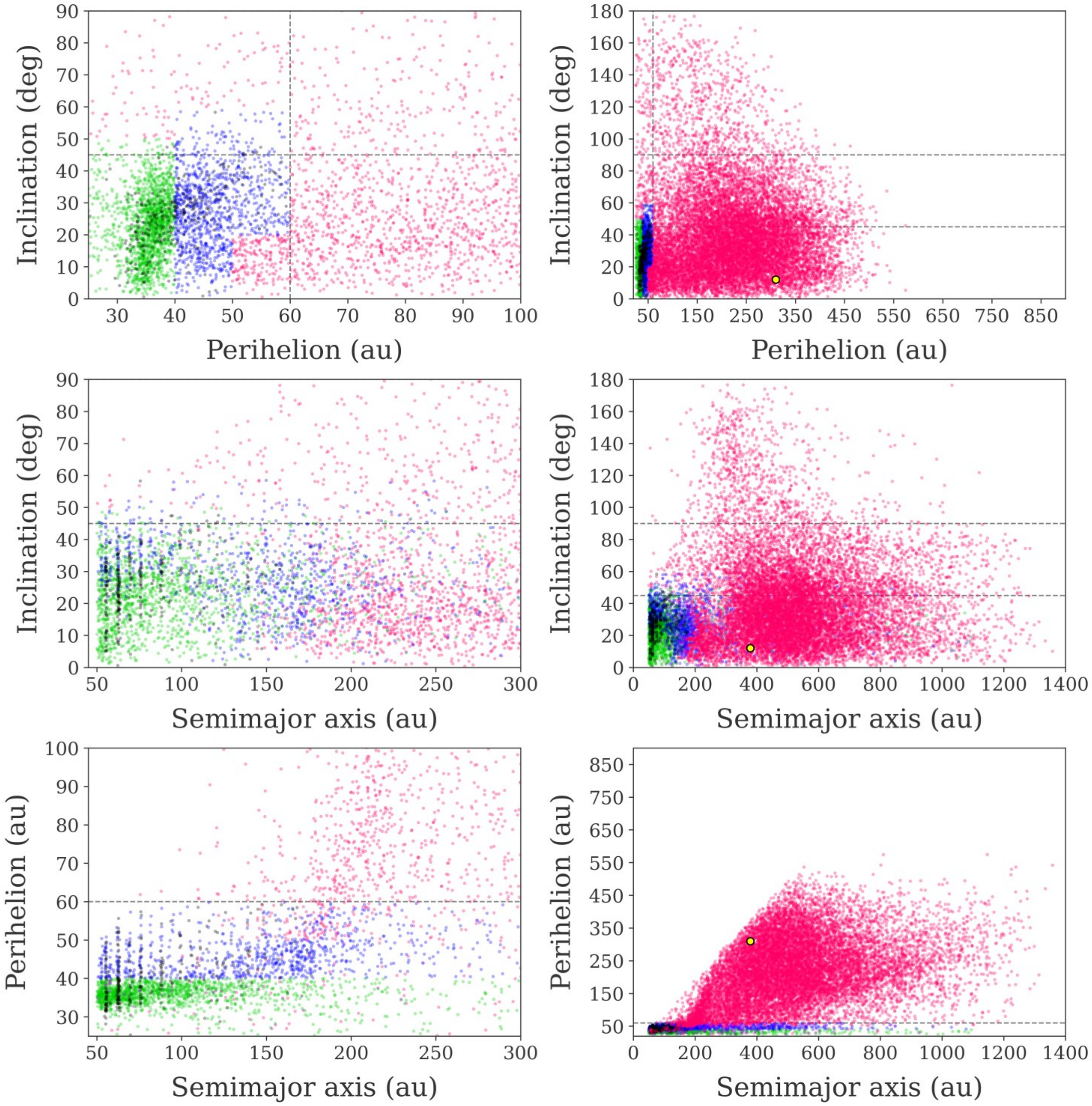

**Figure 15.** Same as Figure 6, but showing the results obtained from the P9 model. In this model, Planet 9 has $a$ = 380 au, $i$ = 16°, $q$ = 300 au, and $m$ = 6.2 $M_{\oplus}$.

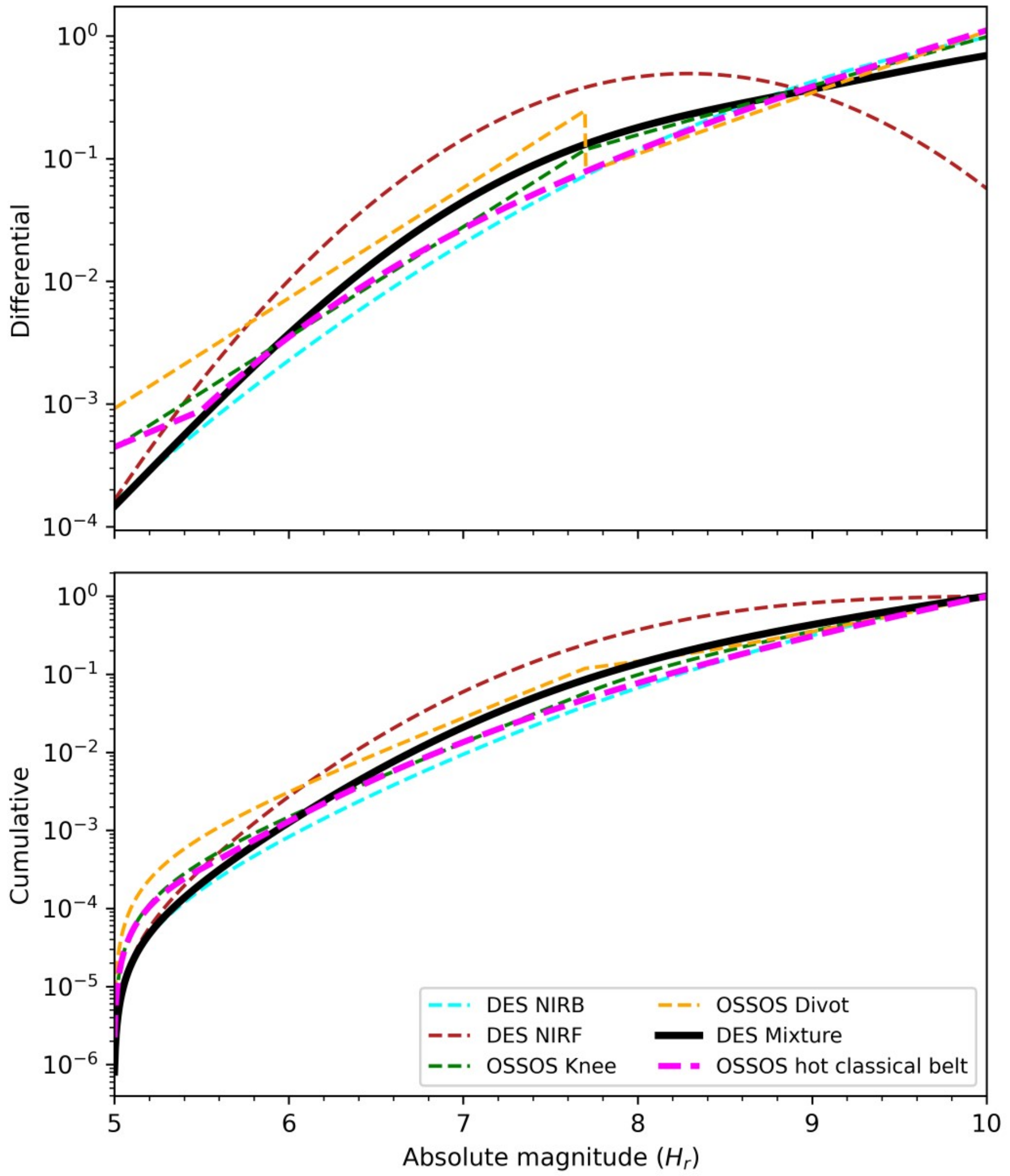

**Figure 16.** The absolute magnitude (*H*) distribution adopted for the synthetic model populations used in the DES Survey Simulator is indicated by the solid black curve, representing a weighted combination of the 'red' DES NIRF (30%) and 'blue' DES NIRB (70%) distributions (Bernardinelli et al. 2025). For comparison, the *H*-distributions constrained by OSSOS observations are also shown: the representative 'divot' and 'knee' distributions (Shankman et al. 2013, 2016; Lawler et al. 2018b) and the hot classical belt distribution (Petit et al. 2023). See Section 4 for further details.

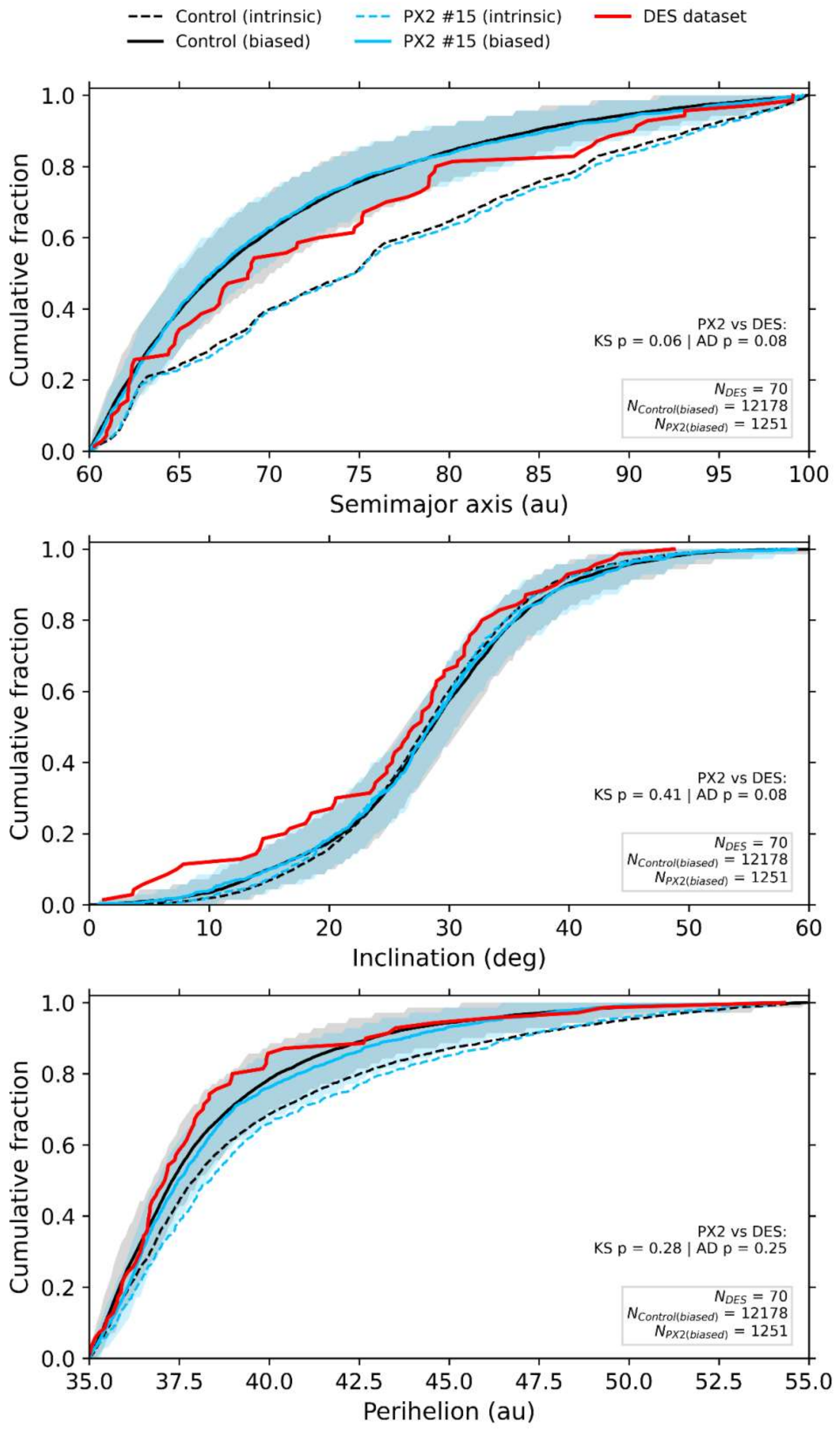

**Figure 17.** Comparison of cumulative orbital distributions for the $a$ = 60–100 au region. The observationally biased results from the control model and the PX2 model 15 ($a$ = 500 au, $i$ = 15°, $q$ = 200 au, and $m$ = 2 $M_\oplus$; see Figure 10) are shown as solid black and dark cyan curves, respectively, while DES observations are shown as red curves. Dashed curves indicate the intrinsic, unbiased distributions for both models. Only detected objects with $q$ = 35–55 au (representing stable distant TNO populations) are included. Sample sizes for the DES data and simulated detections, along with the KS and AD statistical test results, are indicated in each panel. Shaded bands represent the 95% confidence envelopes for the biased models. See Section 4.1 for further details.

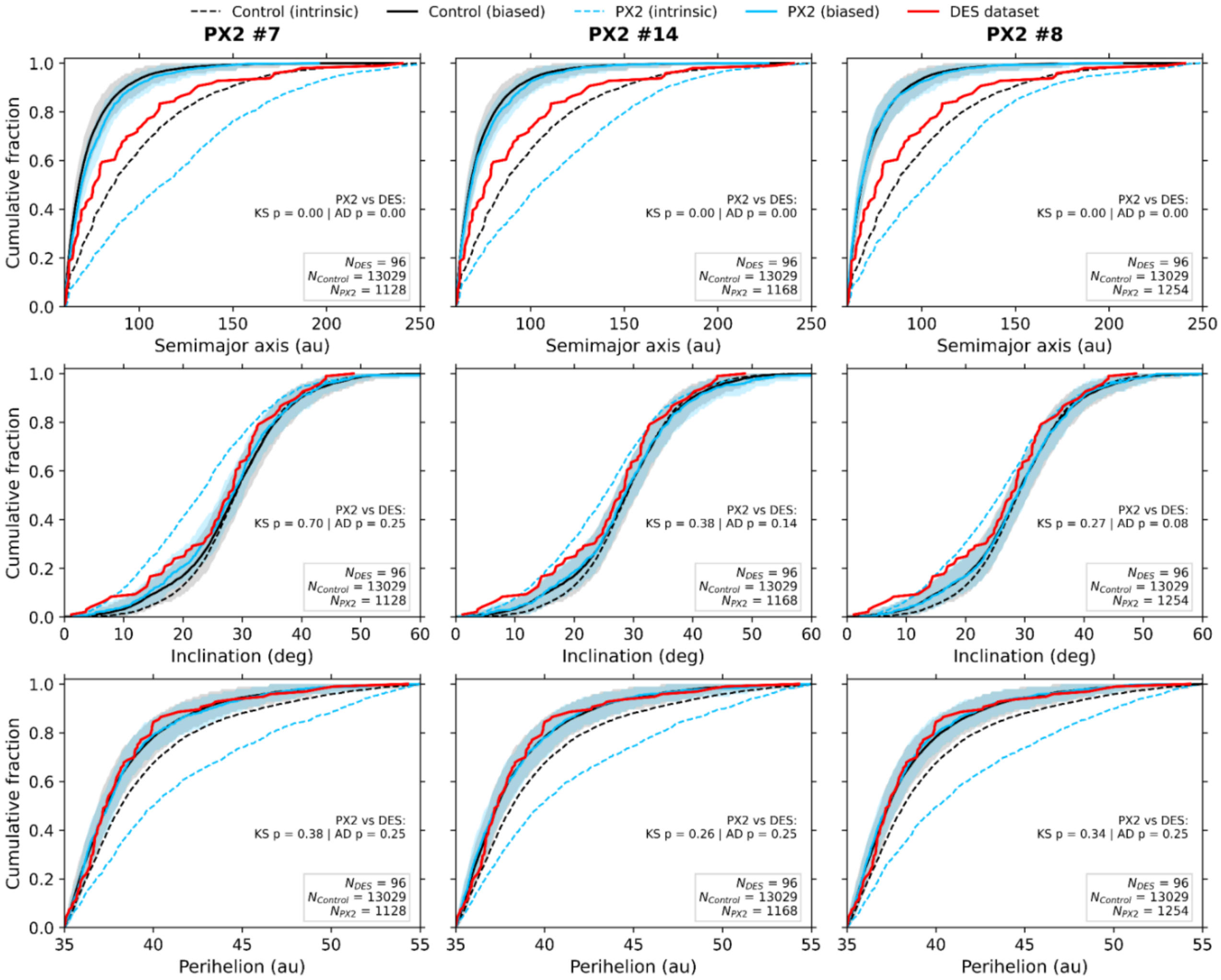

**Figure 18.** Same as Figure 17, but for the $a$ = 60–250 au region. The left, middle, and right three-panel columns show results for three PX2 models with $a$ = 350 au, $q$ = 200 au, and $m$ = 2 $M_{\oplus}$. These columns correspond to Planet X2 inclinations of 1°, 15°, and 30°, respectively (PX2 models 7, 14, and 8; Figures 6–8. See Section 4.1 for further details.

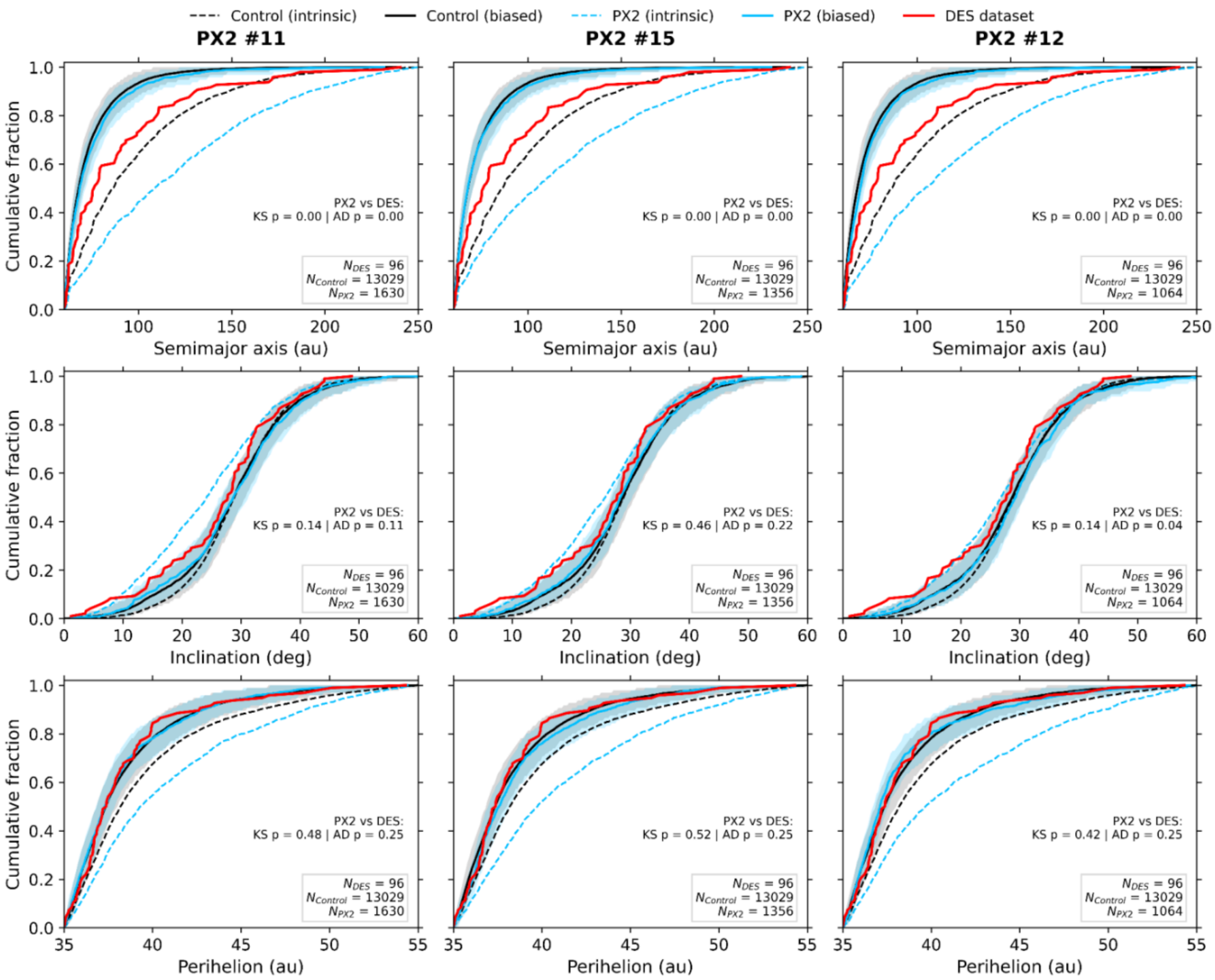

**Figure 19.** Same as Figure 17, but for the $a$ = 60–250 au region. The left, middle, and right three-panel columns show results for three PX2 models with $a$ = 500 au, $q$ = 200 au, and $m$ = 2 $M_\oplus$. These columns correspond to Planet X2 inclinations of 1°, 15°, and 30°, respectively (PX2 models 11, 15, and 12; Figures 9–11. See Section 4.1 for further details.

**Table 1.** Simulation details for the models investigated in this work.

| Model | $m_P$ ($M_\oplus$) | $a_P$ (au) | $q_P$ (au) | $i_P$ (°) | $N_P$ | Interior disk ($a$ = 25–30 au) | | Exterior disk ($a$ = 30–49 au) | Runs | Selected runs |
|---|---|---|---|---|---|---|---|---|---|---|
| | | | | | | Massive component | Massless component | | | |
| Emb-1M3 | 0.3 | 27 | 26.7 | 0.5 | 1 | 1500P+13500 [14.0] | 137000 (87000) | 2500 (1800) | 20 (20) | 31 |
| Emb-1M5 | 0.5 | 27 | 26.7 | 0.5 | 1 | 1500P+13500 [14.0] | 137000 (87000) | 2500 (1800) | 20 (20) | 21 |
| Emb-1E | 1.0 | 27 | 26.7 | 0.5 | 1 | 1500P+13500 [14.0] | 137000 (87000) | 2500 (1800) | 20 (20) | 20 |
| Emb-10M | 0.1 | 25–30 | 24.7-29.7 | 0.5 | 10 | 1500P+13500 [14.0] | 87000 | 2500 | 40 | 16 |
| Emb-2E | 1.0 | 27, 29 | 26.7, 28.7 | 0.5 | 2 | 1500P+13500 [14.0] | 87000 | 2500 | 40 | 13 |
| PX2 #1 | 1.0 | 250 | 200 | 1 | 1 | 1500P+14250 [14.5] | 136000 | 2500 | 10 | 6 |
| PX2 #2 | 1.0 | 250 | 200 | 30 | 1 | 1500P+14250 [14.5] | 136000 | 2500 | 10 | 8 |
| PX2 #3 | 2.0 | 250 | 200 | 1 | 1 | 1500P+14250 [14.5] | 136000 | 2500 | 10 | 8 |
| PX2 #4 | 2.0 | 250 | 200 | 30 | 1 | 1500P+14250 [14.5] | 136000 | 2500 | 10 | 10 |
| PX2 #5 | 1.0 | 350 | 200 | 1 | 1 | 1500P+14250 [14.5] | 136000 | 2500 | 10 | 9 |
| PX2 #6 | 1.0 | 350 | 200 | 30 | 1 | 1500P+14250 [14.5] | 136000 | 2500 | 10 | 9 |
| PX2 #7 | 2.0 | 350 | 200 | 1 | 1 | 1500P+14250 [14.5] | 136000 | 2500 | 10 | 7 |
| PX2 #8 | 2.0 | 350 | 200 | 30 | 1 | 1500P+14250 [14.5] | 136000 | 2500 | 10 | 8 |
| PX2 #9 | 1.0 | 500 | 200 | 1 | 1 | 1500P+14250 [14.5] | 136000 | 2500 | 10 | 9 |
| PX2 #10 | 1.0 | 500 | 200 | 30 | 1 | 1500P+14250 [14.5] | 136000 | 2500 | 10 | 8 |
| PX2 #11 | 2.0 | 500 | 200 | 1 | 1 | 1500P+14250 [14.5] | 136000 | 2500 | 10 | 9 |
| PX2 #12 | 2.0 | 500 | 200 | 30 | 1 | 1500P+14250 [14.5] | 136000 | 2500 | 10 | 8 |
| PX2 #13 | 2.0 | 250 | 200 | 15 | 1 | 1500P+14250 [14.5] | 136000 | 2500 | 10 | 9 |
| PX2 #14 | 2.0 | 350 | 200 | 15 | 1 | 1500P+14250 [14.5] | 136000 | 2500 | 10 | 9 |
| PX2 #15 | 2.0 | 500 | 200 | 15 | 1 | 1500P+14250 [14.5] | 136000 | 2500 | 10 | 9 |
| PX2 #16 | 2.0 | 500 | 250 | 1 | 1 | 1500P+14250 [14.5] | 136000 | 2500 | 10 | 8 |
| PX2 #17 | 2.0 | 500 | 250 | 15 | 1 | 1500P+14250 [14.5] | 136000 | 2500 | 10 | 10 |
| PX2 #18 | 2.0 | 500 | 250 | 30 | 1 | 1500P+14250 [14.5] | 136000 | 2500 | 10 | 8 |
| P9 | 6.2 | 380 | 300 | 16 | 1 | 1500P+14250 [14.5] | 136000 | 2500 | 10 | 8 |

**Notes.** All models began with a protoplanetary disk represented by two main parts: the interior disk, initially set with orbits at $a$ = 25–30 au ($e < 0.2$ and $i < 5°$), and the exterior disk, set at $a$ = 30–49 au ($e < 0.01$ and $i < 0.5°$). The number density of objects in the interior and exterior disks followed distributions proportional to $a^0$ (uniform) and $a^{-1}$, respectively. Numbers in square brackets represent the total mass of the massive component in Earth masses ($M_\oplus$), while numbers in parentheses in specific columns refer to additional lower-resolution runs. A total of 1500 Pluto-class objects on dynamically cold orbits ($e < 0.01$ and $i < 0.5°$) were placed uniformly across 25–30 au, contributing 5 $M_\oplus$ to the total mass of the interior disk's massive component. The Emb models

assumed the existence of scattering planets (primordial Mars- to Earth-mass embryos) formed within the interior disk on dynamically cold orbits ($e$ = 0.01, $i$ = 0.5°). In contrast, the Planet X2 (PX2) models considered a hypothetical outer planet on a stable orbit. The number of planetary bodies, masses, and orbital properties for the scattering-planet and Planet X2 models are denoted by $N_P$, $m_P$, $a_P$, $q_P$, and $i_P$. The representative Planet 9 was specifically investigated in the P9 model. Selected runs yielded systems whose properties satisfied the minimum conditions required for analysis (see Section 2 for other details).

**Table 2.** Key system properties at 4.5 Gyr for each investigated model.

| Model | $m_P$ ($M_\oplus$) | $a_P$ (au) | $q_P$ (au) | $i_P$ (°) | $r_{sca/det}$ | $F$hi (%) | $F$hi$_{sca}$ (%) | $F$hi$_{det}$ (%) | $F$i90 (%) | $F$i90$_{sca}$ (%) | $F$i90$_{det}$ (%) | $A$i25 (au) | $FS$i30 (%) | $F$ext (%) |
|---|---|---|---|---|---|---|---|---|---|---|---|---|---|---|
| *Control-L7* | — | — | — | — | 4.2 | 1.2 | 0.4 | 4.5 | 0.0 | 0.0 | 0.0 | — | — | — |
| *Control-GM* | — | — | — | — | 3.7 | 3.0 | 0.9 | 10.0 | 0.0 | 0.0 | 0.0 | 124.4 | — | 0.5 |
| *PX2 #1* | 1.0 | 250 | 200 | 1 | 1.28 | 2.3 | 1.4 | 3.4 | 0.0 | 0.0 | 0.0 | 196.5 | 83.8 | 12.3 |
| *PX2 #2* | 1.0 | 250 | 200 | 30 | 1.13 | 3.9 | 2.1 | 6.6 | 0.0 | 0.0 | 0.0 | 167.0 | 57.5 | 14.8 |
| *PX2 #5* | 1.0 | 350 | 200 | 1 | 0.78 | 2.3 | 1.3 | 3.0 | 0.0 | 0.0 | 0.0 | 357.2 | 88.4 | 33.3 |
| *PX2 #6* | 1.0 | 350 | 200 | 30 | 1.05 | 7.7 | 1.5 | 16.6 | 0.0 | 0.0 | 0.0 | 206.0 | 34.2 | 23.4 |
| *PX2 #9* | 1.0 | 500 | 200 | 1 | 0.74 | 3.2 | 1.8 | 4.7 | 0.0 | 0.0 | 0.0 | 609.1 | 87.6 | 35.1 |
| *PX2 #10* | 1.0 | 500 | 200 | 30 | 0.80 | 14.5 | 1.4 | 28.1 | 1.2 | 0.0 | 2.4 | 394.7 | 31.0 | 32.8 |
| *PX2 #3* | 2.0 | 250 | 200 | 1 | 0.46 | 2.9 | 1.6 | 3.4 | 0.0 | 0.0 | 0.0 | 241.6 | 81.3 | 47.4 |
| *PX2 #13* | 2.0 | 250 | 200 | 15 | 0.47 | 10.6 | 2.4 | 15.3 | 0.0 | 0.0 | 0.0 | 204.5 | 50.6 | 44.9 |
| *PX2 #4* | 2.0 | 250 | 200 | 30 | 0.46 | 19.5 | 3.3 | 28.6 | 0.9 | 0.0 | 1.4 | 182.9 | 26.4 | 43.5 |
| *PX2 #7* | 2.0 | 350 | 200 | 1 | 0.32 | 7.2 | 2.1 | 9.3 | 0.4 | 0.0 | 0.5 | 433.0 | 74.7 | 59.3 |
| *PX2 #14* | 2.0 | 350 | 200 | 15 | 0.27 | 22.2 | 3.4 | 28.4 | 0.8 | 0.0 | 1.0 | 417.1 | 44.7 | 64.0 |
| *PX2 #8* | 2.0 | 350 | 200 | 30 | 0.27 | 38.5 | 6.3 | 50.2 | 6.5 | 0.6 | 8.4 | 309.5 | 20.7 | 64.4 |
| *PX2 #11* | 2.0 | 500 | 200 | 1 | 0.48 | 16.4 | 3.6 | 23.6 | 6.5 | 1.5 | 9.1 | 447.1 | 56.8 | 49.6 |
| *PX2 #15* | 2.0 | 500 | 200 | 15 | 0.35 | 29.9 | 4.5 | 41.2 | 12.8 | 2.5 | 17.3 | 476.5 | 32.3 | 60.7 |
| *PX2 #12* | 2.0 | 500 | 200 | 30 | 0.24 | 45.6 | 7.9 | 56.5 | 21.9 | 4.5 | 27.2 | 503.2 | 20.7 | 67.8 |
| *PX2 #16* | 2.0 | 500 | 250 | 1 | 0.41 | 15.1 | 2.6 | 21.5 | 4.8 | 0.5 | 6.9 | 574.5 | 61.5 | 54.0 |
| *PX2 #17* | 2.0 | 500 | 250 | 15 | 0.38 | 23.9 | 4.6 | 32.8 | 7.7 | 1.4 | 10.7 | 563.0 | 45.9 | 57.5 |
| *PX2 #18* | 2.0 | 500 | 250 | 30 | 0.36 | 32.8 | 5.3 | 44.4 | 9.5 | 1.5 | 12.7 | 547.3 | 28.7 | 60.2 |
| *P9* | 6.2 | 380 | 300 | 16 | 0.11 | 29.2 | 7.4 | 32.1 | 4.9 | 2.1 | 5.3 | 478.4 | 40.8 | 82.4 |
| Intrinsic values | — | — | — | — | $\lesssim 1.0$ | $\gg 2.0$ | $\gg 1.0$ | $\gg 7.0$ | > 0? | ? | ? | $\gtrsim 250$ | > 50 | ? |

**Notes.** The *Control-L7* model is based on the results of the control model from Lykawka & Ito (2023), which considered only the four giant planets on their current orbits and a primordial scattered disk based on the CFEPS-L7 model (Petit et al. 2011, 2017). The *Control-GM* model is the control model adopted in this work, based on the results of grainy migration models explored in Paper 1. The PX2 and P9 models are consistent with those described in Table 1. The protoplanetary disk was assumed to extend up to 47 au. Only objects with $a > 50$ au and $q > 25$ au were included when evaluating the results in this table. $r_{sca/det}$ is the ratio of the scattered to detached populations. $F$hi and $F$i90 represent the fraction of high-$i$ ($i > 45°$) and retrograde ($i > 90°$) objects, respectively. The subscripts 'sca' and 'det' refer to results for the scattered and detached populations. $A$i25 is the median semimajor axis of detached objects with $i < 25°$.

*FS*i30 is the fraction of Sednoids (objects with $a$ = 200–1500 au and $q > 60$ au) with $i < 30°$. *F*ext is the fraction of extreme objects. The intrinsic values for the population ratio $r_{sca/det}$ (e.g., Beaudoin et al. 2023) and high-$i$ fractions reflect the assumed values discussed in Section 1. The intrinsic values of *A*i25 and *FS*i30 were adopted based on the findings of Kaib et al. (2025) and Hu et al. (2025), respectively. All results are median values obtained across the selected runs in each model (Table 1). See the main text for further details.